\documentclass[fleqn,usenatbib,useAMS]{mnras}

\usepackage{graphicx}	
\usepackage{amsmath}	
\usepackage{amssymb}	
\usepackage{multicol}        
\usepackage{bm}		
\usepackage{pdflscape}	
\usepackage{booktabs}  
\usepackage{multirow}
\usepackage{stackengine}

\usepackage{subcaption}
\usepackage{color}   

\definecolor{titlecol4}{rgb}{0.039,0.361,0.569} 

\usepackage[T1]{fontenc}
\usepackage{ae,aecompl}

\usepackage{newtxtext,newtxmath}

\usepackage{hyperref}

\title[Galaxy Zoo DECaLS Data Release]{Galaxy Zoo DECaLS: Detailed Visual Morphology Measurements from Volunteers and Deep Learning for 314,000 Galaxies}

\author[M. Walmsley et al]{Mike Walmsley$^{1}$\thanks{Contact e-mail: \href{mailto:mike.walmsley@physics.ox.ac.uk}{mike.walmsley@physics.ox.ac.uk}},
Chris Lintott$^{1}$,
Tobias G\'eron$^{1}$,
Sandor Kruk$^{2}$,
Coleman Krawczyk$^{3}$,
\newauthor
Kyle W. Willett$^{4}$,
Steven Bamford$^{5}$,
Lee S. Kelvin$^{6}$,
Lucy Fortson$^{7}$,
Yarin Gal$^{8}$,
\newauthor
William Keel$^{9}$,
Karen L. Masters$^{10}$,
Vihang Mehta$^{9}$,
Brooke D. Simmons$^{11}$,
\newauthor
Rebecca Smethurst$^{1}$,
Lewis Smith$^{8}$,
Elisabeth M. Baeten$^{12}$,
Christine Macmillan$^{12}$ 
\\
$^{1}$Oxford Astrophysics, Department of Physics, University of Oxford, Denys Wilkinson Building, Keble Road, Oxford, OX1 3RH, UK\\
$^{2}$European Space Agency, ESTEC, Keplerlaan 1, NL-2201 AZ, Noordwijk, The Netherlands\\
$^{3}$Institute of Cosmology and Gravitation, University of Portsmouth Dennis Sciama Building, Burnaby Road, Portsmouth, PO1 3FX, UK\\
$^{4}$School of Physics and Astronomy, University of Minnesota, 116 Church St SE, Minneapolis, MN 55455, USA\\
$^{5}$School of Physics and Astronomy, University of Nottingham, University Park, Nottingham, NG7 2RD, UK\\
$^{6}$Department of Astrophysical Sciences, Princeton University, 4 Ivy Lane, Princeton, NJ 08544, USA\\
$^{7}$Minnesota Institute for Astrophysics, University of Minnesota, 116 Church St SE, Minneapolis, MN 55455, USA\\
$^{8}$Oxford Applied and Theoretical Machine Learning (OATML) Group, Department of Computer Science, University of Oxford, Oxford, OX1 3QD, UK\\
$^{9}$Dept. of Physics and Astronomy, University of Alabama, Tuscaloosa, AL 35487, USA\\
$^{10}$Department of Physics and Astronomy, Haverford College, 370 Lancaster Avenue, Haverford, PA 19041, USA\\
$^{11}$Department of Physics, Lancaster University, Bailrigg, Lancaster, LA1 4YB, UK\\
$^{12}$Citizen Scientist, Zooniverse c/o University of Oxford, Keble Road, Oxford OX1 3RH, UK
}

\date{Last updated XXX; in original form XXX}

\pubyear{2021}

\begin{document}
\label{firstpage}
\pagerange{\pageref{firstpage}--\pageref{lastpage}}
\maketitle

\begin{abstract}
We present Galaxy Zoo DECaLS: detailed visual morphological classifications for Dark Energy Camera Legacy Survey images of galaxies within the SDSS DR8 footprint. Deeper DECaLS images ($r=23.6$ vs $r=22.2$ from SDSS) reveal spiral arms, weak bars, and tidal features not previously visible in SDSS imaging. To best exploit the greater depth of DECaLS images, volunteers select from a new set of answers designed to improve our sensitivity to mergers and bars. Galaxy Zoo volunteers provide 7.5 million individual classifications over 314,000 galaxies. 140,000 galaxies receive at least 30 classifications, sufficient to accurately measure detailed morphology like bars, and the remainder receive approximately 5. All classifications are used to train an ensemble of Bayesian convolutional neural networks (a state-of-the-art deep learning method) to predict posteriors for the detailed morphology of all 314,000 galaxies. We use active learning to focus our volunteer effort on the galaxies which, if labelled, would be most informative for training our ensemble. When measured against confident volunteer classifications, the trained networks are approximately 99\% accurate on every question. Morphology is a fundamental feature of every galaxy; our human and machine classifications are an accurate and detailed resource for understanding how galaxies evolve.
\end{abstract}




\section{Introduction}

Morphology is a key driver and tracer of galaxy evolution. For example, bars are thought to move gas inwards \citep{Sakamoto1999} driving and/or shutting down star formation \citep{Sheth2004, Jogee2005}, and bulges are linked to global quenching \citep{Masters2011,Fang2013,Bluck2014} and inside-out quenching \citep{Spindler2017,Lin2019}. Morphology also traces other key drivers, such as the merger history of a galaxy. Mergers support galaxy assembly \citep{Wang2011, Martin2018a}, though their relative contribution is an open question \citep{Casteels2014}, and may create tidal features, bulges, and disks, allowing past mergers to be identified \citep{Hopkins2010, Fontanot2011, Kaviraj2014, Brooks2015}.

Unpicking the complex interplay between morphology and galaxy evolution requires measurements of detailed morphology in large samples. While modern surveys reveal exquisite morphological detail, they image far more galaxies than scientists can visually classify. Galaxy Zoo solves this problem by asking members of the public to volunteer as `citizen scientists' and provide classifications through a web interface. Galaxy Zoo has provided morphology measurements for surveys including SDSS \citep{Lintott2008, Willett2013} and large HST programs \citep{Simmons2017, Willett2017a}.

Knowing the morphology of homogeneous samples of hundreds of thousands of galaxies supports science only possible at scale. The catalogues produced by the collective effort of Galaxy Zoo volunteers have been used as the foundation of a large number of studies of galaxy morphology (see \citealt{Masters2019a} for a review), with the method's ability to provide estimates of confidence alongside classification especially valuable.  Galaxy Zoo measures subtle effects in large populations \citep{Masters2010, Willett2015, Hart2017Stars}; identifies unusual populations that challenge standard astrophysics \citep{Simmons2013,Tojeiro2013,Kruk2017}; and finds unexpected and interesting objects that provide unique data on broader galaxy evolution questions \citep{Lintott2009,Cardamone2009,Keel2015}.

Here, we present the first volunteer classifications of galaxy images collected by the Dark Energy Camera Legacy Survey (DECaLS, \citealt{Dey2018}). This work represents the first systematic engagement of volunteers with low-redshift images as deep as those provided by DECaLS, and thus represents a more reliable catalogue of detailed morphology than has hitherto been available. These detailed classifications include the presence and strength of bars and bulges, the count and winding of spiral arms, and the indications of recent or ongoing mergers. Our volunteer classifications were sourced over three separate Galaxy Zoo DECaLS (GZD) classification campaigns, GZD-1, GZD-2, and GZD-5, which classified galaxies first released in DECaLS Data Releases 1, 2, and 5 respectively. The key practical differences are that GZD-5 uses an improved decision tree aimed at better identification of mergers and weak bars, and includes galaxies with just 5 total votes as well as galaxies with 40 or more. Across all campaigns, we collect 7,496,325 responses from Galaxy Zoo volunteers, recording 30 or more classifications in at least one campaign for 139,919 galaxies and fewer (approximately 5 classifications) for an additional 173,870 galaxies, totalling 313,789 classified galaxies.

For the first time in a Galaxy Zoo data release, we also provide automated classifications made using Bayesian deep learning \citep{Walmsley2020}. By using our volunteer classifications to train a deep learning algorithm, we can make detailed classifications for all 313,789 galaxies in our target sample, providing morphology measurements faster than would be possible than relying on volunteers alone. Bayesian deep learning allows us to learn from uncertain volunteer responses and to estimate the uncertainty of our predictions. It also allows us to identify which galaxies, if labelled, would be most informative for training our classifier (active learning). We chose to partially focus our volunteers on such informative galaxies, requesting 40 classifications per informative galaxy and only 5 for the remainder. Our classifier predicts posteriors for how volunteers would have answered all decision tree questions\footnote{Excluding the final `Is there anything odd?' question as it is multiple-choice}, with an accuracy comparable to asking 5 to 15 volunteers, depending on the question, and achieving approximately 99\% accuracy on every question for galaxies where the volunteers are confident (volunteer vote fractions below 0.2 or above 0.8).

In Section \ref{imaging}, we describe the observations used and the creation of RGB images suitable for classification. 
In Section \ref{volunteer_classifications}, we give an overview of the volunteer classification process and detail the new decision trees used.
In Section \ref{sec:volunteer_analysis}, we investigate the effects of improved imaging and improved decision trees, and we compare our results to other morphological measurements.
Then, in Section \ref{sec:automated}, we describe the design and performance of our automated classifier - an ensemble of Bayesian convolutional neural networks. Finally, in Section \ref{sec:usage}, we provide guidance (and example code) for effective use of the classifications.

\section{Imaging}
\label{imaging}

\subsection{Observations}

Our galaxy images are created from data collected by the DECaLS survey \citep{Dey2018}. 
DECaLS uses the Dark Energy Camera (DECam, \citealt{Flaugher2015}) at the 4m Blanco telescope at Cerro Tololo Inter-American Observatory, near La Serena, Chile.  DECam has a roughly hexagonal 3.2 square degree field of view with a pixel scale of 0.262 arcsec per pixel. The median point spread function FWHM is $1 \farcs 29$, $1 \farcs 18$ and $1\farcs 11$ for $g$, $r$, and $z$, respectively.

The DECaLS survey contributes targeting images for the upcoming Dark Energy Spectroscopic Instrument (DESI). DECaLS is responsible for the DESI footprint in the Southern Galactic Cap (SGC) and the $\delta \leq 34$ region of the Northern Galactic Cap (NGC), totalling 10,480 square degrees\footnote{The remaining DESI footprint is being imaged by DECaLS' companion surveys, MzLS and BASS \citep{Dey2018}}. 1130 square degrees of the SGC DESI footprint are already being imaged by DECam through the Dark Energy Survey (DES, \citealt{TheDarkEnergySurveyCollaboration2005}) so DECaLS does not repeat this part of the DESI footprint.
DECaLS implements a 3-pass strategy to tile the sky. Each pass is slightly offset (approx 0.1-0.6 $\deg$). The choice of pass and exposure time for each observation is optimised in real-time based on the observing conditions recorded for the previous targets, as well as the interstellar dust reddening, sky position, and estimated observing conditions of possible next targets. This allows a near-uniform depth across the survey. In DECaLS DR1, DR2, and DR5, from which our images are drawn, the median $5\sigma$ point source depths for areas with 3 observations was approximately (AB) $g$=24.65, $r$=23.61, and $z$=22.84\footnote{See https://www.legacysurvey.org/dr5/description/ and related pages}. The DECaLS survey completed observations in March 2019.

\subsection{Selection}
\label{sec:selection}

\begin{figure}
    \centering
    \includegraphics[width=\columnwidth]{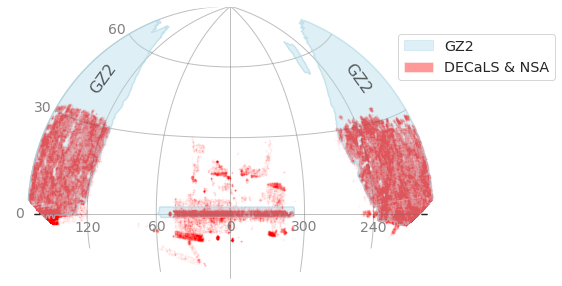}
    \caption{Sky coverage of GZ DECaLS (equatorial coordinates), resulting from the imaging overlap of DECaLS DR5 and SDSS DR8, shown in red. Darker areas indicate more galaxies. Sky coverage of Galaxy Zoo 2, which used images sourced from SDSS DR7, shown in light blue. The NSA includes galaxies imaged by SDSS DR8, including galaxies newly imaged at the Southern Galactic Cap (approx. $2500\deg^2$)}
    \label{fig:footprints}
\end{figure}

We identify galaxies in the DECaLS imaging using the NASA-Sloan Atlas v1.0.0 (NSA). As the NSA was derived from SDSS DR8 imaging \citep{Aihara2011}, this data release only includes galaxies that are within both the DECaLS and SDSS DR8 footprint. In effect, we are using deeper DECaLS imaging of the galaxies previously imaged in SDSS DR8. This ensures our morphological measurements have a wealth of ancillary information derived from SDSS and related surveys, and allows us to measure any shift in classifications vs. Galaxy Zoo 2 using the subset of SDSS DR8 galaxies classified both in this work and in Galaxy Zoo 2 (Sec. \ref{sec:volunteer_analysis}). Figure \ref{fig:footprints} shows the resulting GZ DECaLS sky coverage. NSA v1.0.0 was not published but the values of the columns used here are identical to those in NSA v1.0.1, released in SDSS DR13 \citep{Albareti2017}; only the column naming conventions are different.

Selecting galaxies with the NSA introduces two implicit cuts.
First, the NSA primarily includes galaxies brighter than $m_r=17.77$, the SDSS spectroscopic target selection limit. Galaxies fainter than $m_r=17.77$ are included only if they are in deeper survey areas (e.g. Stripe82) or were measured using `spare' fibres after all brighter galaxies in a given field were covered; we suggest researchers enforce their own magnitude cut according to their science case. Second, the NSA only covers redshifts of $z=0.15$ or below. To these implicit cuts, we add an explicit cut requiring Petrosian radius (the NSA v1.0.0 \texttt{PETROTHETA}\footnote{Azimuthally-averaged SDSS-style Petrosian radius, derived from the r band. See \cite{Albareti2017} and the \href{https://data.sdss.org/datamodel/files/ATLAS_DATA/ATLAS_MAJOR_VERSION/nsa.html}{NSA v1.0.1 data model}.} column) of at least 3 arcseconds, to ensure the galaxy is sufficiently extended for meaningful classification.

For each galaxy, if the coordinates had been imaged in the $g$, $r$ and $z$ bands, and the galaxy passed the selection cuts above, we acquired a combined FITS cutout of the $grz$ bands from the DECaLS cutout service (www.legacysurvey.org). 

Galaxy Zoo presents volunteers with  $424 \times 424$ pixel square 
galaxy images. GZD-1 and GZD-2 acquired  $424 \times 424$ pixel square FITS cutouts directly from the cutout service. To ensure that galaxies typically fit well within a 424 pixel image, cutouts were downloaded with an interpolated pixel scale $s$ of
\begin{equation}
    s = \max(\min(0.04 p_{50}, 0.02 p_{90}), 0.1)
\end{equation}
where $p_{50}$ is the Petrosian 50\%-light radius and $p_{90}$ is the Petrosian 90\%-light radius. Approximately 1\% of galaxies have incorrectly large radii reported in the NSA (typically as a result of foreground stars or other interloping sources) and this causes the field to be incorrectly large and hence the target galaxy to appear incorrectly small. To allow researchers to mitigate this issue, we flag images for which there are more source pixels away from the centre than near the centre; specifically, for which the mean distance of all likely source pixels\footnote{Arbitrarily defined as pixels with double the 20th percentile band-averaged value after the scaling in Sec. \ref{sec:image_construction}.} exceeds 161 (approximately the expected value for all pixels). We find by eye that this simple procedure identifies the worst-affected galaxies. We report the mean source pixel distance and distance flags as \texttt{wrong\_size\_statistic} and \texttt{wrong\_size\_warning}, respectively.

For GZD-5, to avoid banding artifacts caused by the interpolation method of the DECaLS cutout service, each FITS image was downloaded at the fixed native telescope resolution of 0.262 arcsec$^2$ per pixel\footnote{Up to a maximum of 512 pixels per side. Highly extended galaxies were downloaded at reduced resolution such that the FITS had exactly 512 pixels per side.}, with enough pixels to cover the same area as 424 pixels at the interpolated pixel scale $s$. These individually-sized FITS were then resized locally up to the interpolated pixel scale $s$ by Lanczos interpolation \citep{Lanczos1938}. Image processing is otherwise identical between campaigns. Galaxies with incomplete imaging, defined as more than 20\% missing pixels in any band, were discarded. For GZD-1/2, 92,960 of 101,252 galaxies had complete imaging (91.8\%). For GZD-5, 216,106 of 247,746 galaxies not in DECaLS DR1/2 had complete imaging (87.2\%)\footnote{Note that these numbers do not sum to the total number of galaxies classified across both campaigns because some galaxies are shared between campaigns.}.

\subsection{RGB Image Construction}
\label{sec:image_construction}

We convert the measured \textit{grz} fluxes into RGB images following the methodology of \cite{Lupton2004}. To use the \textit{grz} bands as RGB colours, we multiply the flux values in each band by 125.0, 71.43, and 52.63, respectively. These numbers are chosen by eye\footnote{By Dustin Lang, who we gratefully acknowledge.} such that typical subjects show an appropriate range of color once mapped to RGB channels.

For background pixels with very low flux, and therefore high variance in the proportion of flux per band, naively colouring by the measured flux creates a speckled effect \citep{Willett2017a}. As an extreme example, a pixel with 1 photon in the $g$ band and no photons in $r$ or $z$ would be rendered entirely red. To remove these colourful speckles, we desaturate pixels with very low flux. We first estimate the total per-pixel photon count $N$ assuming an exposure time of 90 seconds per band and a mean photon frequency of 600nm. Poisson statistics imply the standard deviation on the total mean flux in that pixel is proportional to $\sqrt{N}$. For pixels with a standard deviation below 100, we scale the per-band deviation from the mean per-pixel flux by a factor of 1\% of the standard deviation. The effect is to reduce the saturation of low-flux pixels in proportion to the standard deviation of the total flux. Mathematically, we set
\begin{equation}
    X^{\prime}_{ijc} = \overline{X_{ij}} + \alpha (X_{ijc} - \overline{X_{ij}}) \quad \text{where} \quad \alpha = \min(0.01 \sqrt{\overline{X_{ij}}T / \lambda}, 1) \quad
\end{equation}
where $X_{ijc}$ and $X^{\prime}_{ijc}$ are the flux at pixel $ij$ in channel $c$ before and after desaturation, $\overline{X_{ij}}$ is the mean flux across bands at pixel $ij$, $T$ is the mean exposure time (here, 90 seconds) and $\lambda$ is the mean photon wavelength (here, 600\,nm).

Pixel values were scaled by $\text{arcsinh}(x)$ to compensate for the high dynamic range typically found in galaxy flux, creating images which can show both bright cores and faint outer features. To remove the very brightest and darkest pixels, we linearly rescale the pixel values to lie on the $(-0.5,\ 300)$ interval and then clip the pixel values to 0 and 255 respectively. We use these final values to create an RGB image using \texttt{pillow} \citep{Kemenade2020}.

The images are available on Zenodo at \href{https://doi.org/10.5281/zenodo.4573248}{https://doi.org/10.5281/zenodo.4573248}.
The code used to download the FITS cutouts and convert them to RGB images is available on GitHub for \href{https://github.com/willettk/decals/blob/b55170aadbfd6ceccd078d6119821db00311e9dd/python/decals.py}{GZD-1},  \href{https://github.com/willettk/decals/blob/master/python/decals_dr2.py}{GZD-2} and \href{https://github.com/zooniverse/decals/blob/master/decals/a_download_decals/get_images/download_images_threaded.py}{GZD-5}.

\section{Volunteer Classifications}
\label{volunteer_classifications}

Volunteer classifications for GZ DECaLS were collected during three campaigns. GZD-1 and GZD-2 classified all 99,109 galaxies passing the criteria above from DECALS DR1 and DR2, respectively. GZD-1 ran from September 2015 to February 2016, and GZD-2 from April 2016 to February 2017. GZD-5 classified 262,000 DECALS DR5-only galaxies passing the criteria above. GZD-5 ran from March 2017 to October 2020. GZD-5 used more complex retirement criteria aimed at improving our automated classification (\ref{sec:retirement}) and an improved decision tree aimed at better identification of weak bars and minor mergers (\ref{sec:comparison_of_decision_trees}).

This iteration of the Galaxy Zoo project used the infrastructure made available by the Zooniverse platform; in particular, the \href{https://github.com/zooniverse/panoptes}{open source Panoptes platform} \citep{TheZooniverseTeam2020}. The platform allows for the rapid creation of citizen science projects, and presents participating volunteers with one of a subject set of images chosen either randomly, or through criteria described in section \ref{sec:retirement}.

\subsection{Selecting Total Classifications}
\label{sec:retirement}

\begin{figure}
    \centering
    \includegraphics[width=\columnwidth]{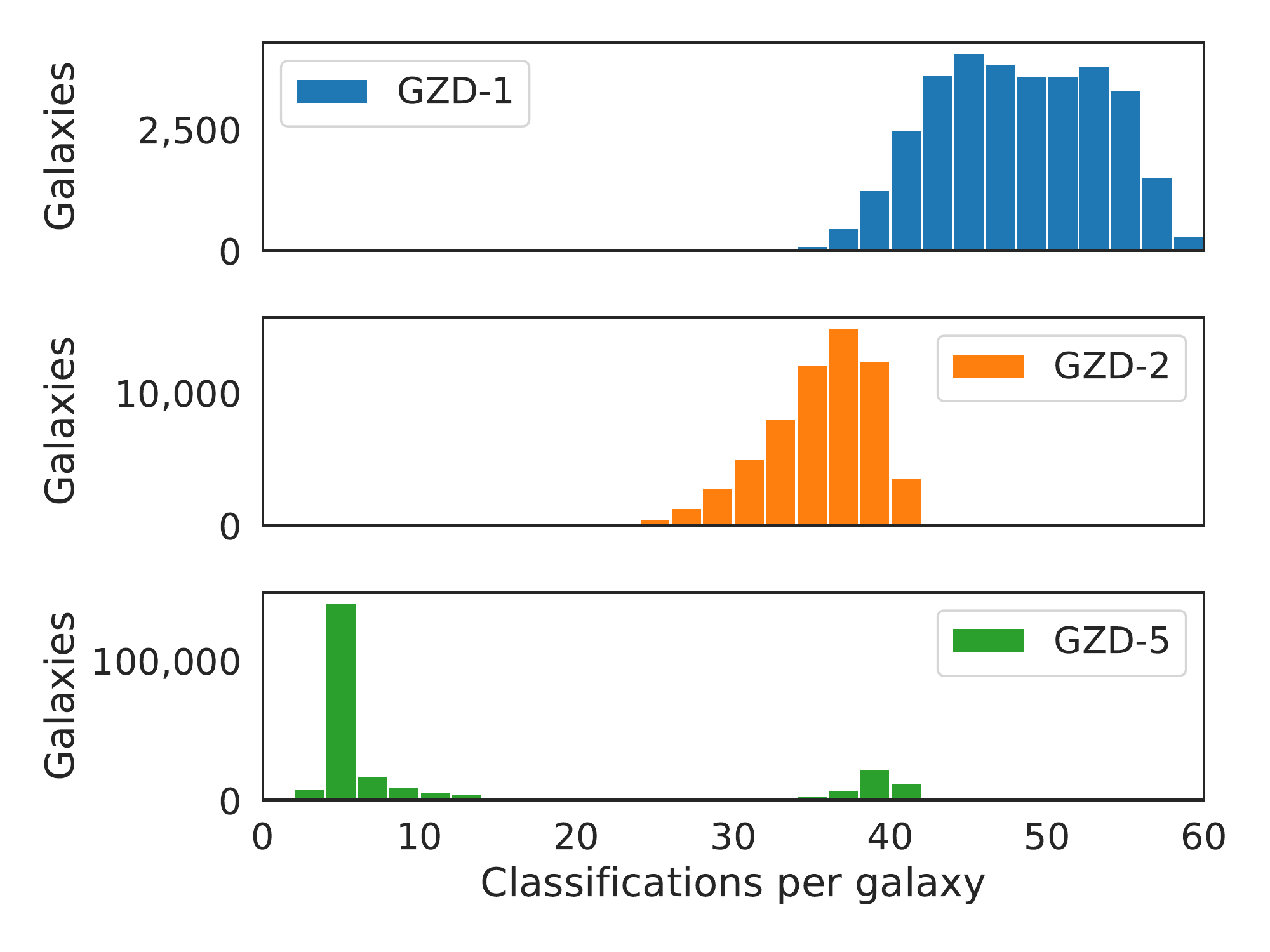}
    \caption{GZD-1, GZD-2 and GZD-5 classification counts, excluding implausible classifications (Sec. \ref{sec:volunteer_weighting}). GZD-1 has approximately 40-60 classifications, GZD-2 has approximately 40, and GZD-5 has either approximately 5 or approximately 30-40. 5.9\% of GZD-5 galaxies received more than 40 classifications due to mistaken duplicate uploads.}
    \label{fig:classification_counts}
\end{figure}

How many volunteer classifications should each galaxy receive? Ideally, all galaxies would receive enough classifications to be confident in the average response (i.e. the vote fraction) while still classifying all the target galaxies within a reasonable timeframe. However, the size of modern surveys make this increasingly impractical. Collecting 40 volunteer classifications for all 314,000 galaxies in this data release would have taken around eight years without further promotion efforts. The larger data sets of future surveys will only be more challenging. In anticipation of future classification demands, we have therefore implemented a variable retirement rate here \citep[motivated and described further in][]{Walmsley2020}. Unlike previous data releases, GZ DECaLS galaxies each received different numbers of classifications (Figure \ref{fig:classification_counts}). Beginning part-way through GZD-5, we prioritise classifications for the galaxies expected to most improve our machine learning models, and rely more heavily on those models for classifying the remainder.  

For GZD-1 and GZD-2, all galaxies received at least 40 classifications\footnote{Note that because classifications from volunteers who respond `artifact' at implausibly high rates are discounted, the total classifications in Fig. \ref{fig:classification_counts} and the published catalog are slightly lower - see Sec. \ref{sec:volunteer_weighting}.} (as with previous data releases). GZD-1 galaxies have between 40 and 60 classifications, selected at random, while GZD-2 galaxies all have approximately 40. For GZD-5, galaxies classified until June 2019 also received approximately 40 classifications. From June 2019, we introduced an active learning system. Using active learning, galaxies expected to be the most informative for training our deep learning model received 40 classifications, and the remaining galaxies received at least 5 classifications.

By `most informative', we mean the galaxies which, if classified, would most improve the performance of our model. We describe our method for estimating which galaxies would be most informative in full in \cite{Walmsley2020}. Briefly, we use a convolutional neural network to make repeated predictions for the probability that $k$ of $N$ total volunteers select `Featured' to the `Smooth or Featured' question\footnote{`Artifact' answers are sufficiently rare that we chose to ignore votes for this answer when calculating which galaxies to label.}. For each prediction, we randomly permute the network with MC Dropout \citep{Gal2016Uncertainty}, approximating (roughly) training many networks to make predictions on the same dataset. It can be shown that, under some assumptions, the most informative galaxies will be the galaxies with confidently different predictions under each MC Dropout permutation; that is, where the permuted networks confidently disagree \citep{Houlsby2014}. Formally, we acquire (label with volunteers) the galaxies with the highest estimated mutual information, given by:
\begin{equation}
    \label{eqn:acquisition}
    \begin{split}
        \mathbb{I}[k, w] & = \\
    & - \sum_{k=0}^{N} \langle\text{Bin}(k|f^w(x), N)\rangle \log[ \langle\text{Bin}(k|f^w(x), N)\rangle] \\
    & + \langle \sum_{k=0}^{N} \text{Bin}(k|f^w(x), N) \log[ \text{Bin}(k|f^w(x), N)] \rangle
    \end{split}
\end{equation}

where $f^w(x)$ is the output of the neural network trained to predict the typical volunteer response following \cite{Walmsley2020} and $\text{Bin}(k|f^w(x), N)$ is the probability for $k$ of $N$ volunteers to answer `Featured' to `Smooth or Featured' given that network-estimated typical response. Angled brackets indicate the expectation over the distribution of weights, approximated as the expectation over MC Dropout permutations. In short, the negative term gives the entropy of the volunteer vote distribution given the mean model predictions, and the positive term gives the mean entropy from the predictions of each permuted model. The difference between these terms measures the degree of confident disagreement between permuted models. See \cite{Walmsley2020} for more.

We used the same architecture and loss function as in \cite{Walmsley2020} while concurrently developing the more sophisticated classifier introduced in this Section. The initial training set was all GZD-5 galaxies fully classified ($N > 36$) by the time of activation. Each active learning cycle proceeded as follows. The model was retrained with all galaxies fully classified by the cycle start date. Next, unlabelled galaxies were ranked by mutual information (Eqn. \ref{eqn:acquisition}) and the most informative 1000 of a random 32768\footnote{To allow for out-of-memory shuffling, binary-encoded galaxy images were stored in `shards' of 4096 galaxies each. 32,768 corresponds to 8 such shards} galaxies were uploaded. Once those galaxies were fully classified by volunteers (typically in 1-4 weeks) the cycle was repeated. 6,939 total galaxies were uploaded in total\footnote{Technical errors with duplicate uploads led to some active-learning-prioritised galaxies receiving more than 40 classifications; the median number of classifications is 44.}.

We chose to select from a subset of galaxies not yet classified for two reasons. The first was for computational efficiency: calculating the acquisition function requires making 5 predictions per galaxy.
The second was that ad hoc experiments showed that galaxies with the very highest acquisition function values were often highly unusual and might be \textit{too} unusual to learn from effectively.
We also added a retirement rule to retire galaxies receiving 5 classifications of `artifact', to help avoid volunteers being presented with these prioritised artifacts.

\begin{figure}
    \centering
    \includegraphics[width=\columnwidth]{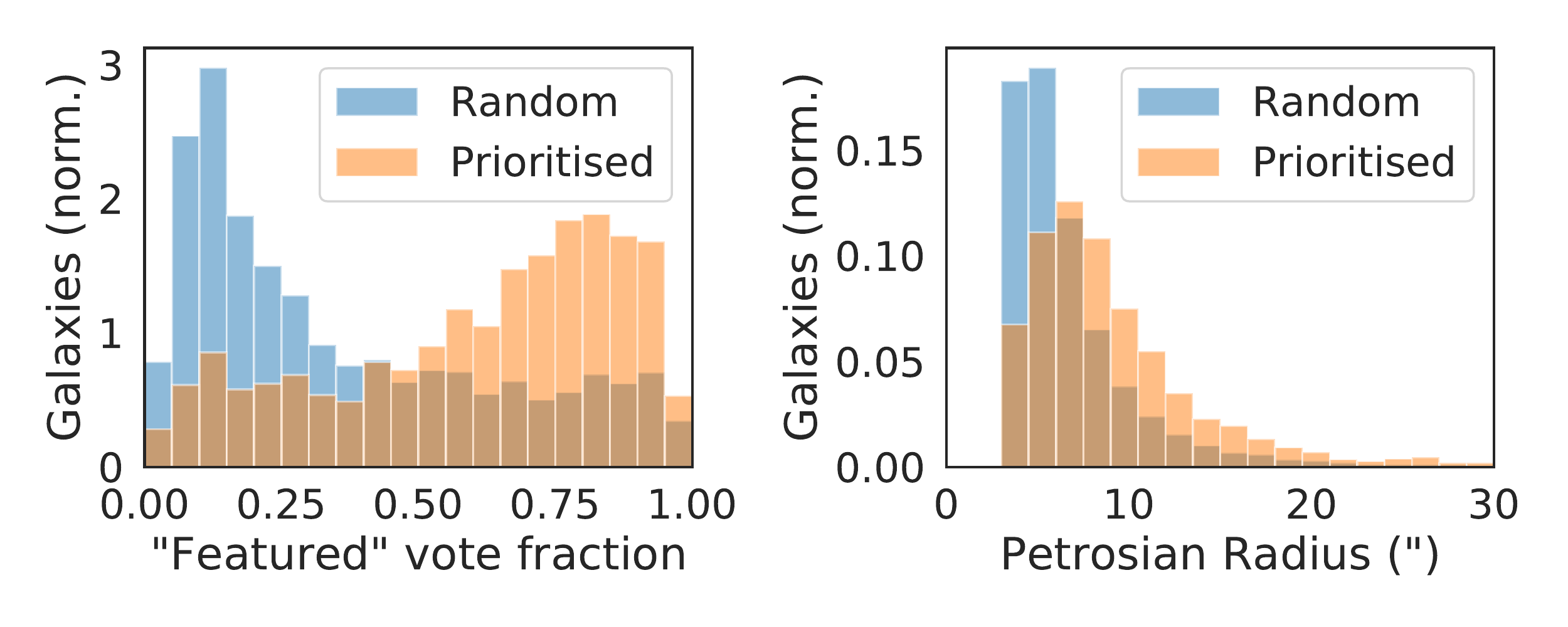}
    \caption{`Featured' vote fraction and Petrosian radius (as measured by the NSA \texttt{PETROTHETA} column) for galaxies selected either at random (prior to enabling active learning) or prioritised as informative. Prioritised galaxies are dramatically more featured and slightly more extended than the previously-classified random galaxies.}
    \label{fig:active_learning_metadata}
\end{figure}

We emphasise that the number of classifications each galaxy received under active learning \textit{is not random}. Figure \ref{fig:active_learning_metadata} shows how active-learning-prioritised galaxies are dramatically more featured and slightly more extended than the previously-classified random galaxies, matching our intuition that small `smooth' elliptical galaxies are easier to classify and hence less informative than extended `featured' galaxies. For details on handling this and other selection effects, see Sec. \ref{sec:usage}.

\subsection{Decision Trees}
\label{sec:decision_trees_intro}

The questions and answers we ask our volunteers define the measurements we can publish. It is therefore critical that the Galaxy Zoo decision tree matches the science goals of the research community. 

The questions in a given Galaxy Zoo workflow are designed to be answerable even by a classifier with little or no astrophysical background. This motivates a focus primarily on the appearance of the galaxy, rather than incorporating physical interpretations which would require prior knowledge of galaxies. As an example, the initial question in all decision trees from Galaxy Zoo 2 onwards has asked the viewer to distinguish primarily between ``smooth'' and ``featured'' galaxies, rather than ``elliptical'' and ``disk'' galaxies. This distinction between descriptive and interpretive classification is not always perfectly enforced. For example, the ``features'' response to the initial question is worded as ``features or disk'', and a later question asks whether the galaxy is ``merging or disturbed'', which requires some interpretation\footnote{The step from visual description to interpretation may explain why a model trained by \cite{Fischer2018} on expert T-Type labels makes more confident predictions than volunteers on whether a subset of low-mass GZ2 galaxies show spiral structure; see \cite{Peterken2021}.}. To aid classifiers, all iterations of Galaxy Zoo have therefore included illustrative icons in the classification interface. Additional help is also available; in the current project, the interface includes a brief tutorial, a detailed field guide with multiple examples of each type of galaxy, and specific help text available for each individual classification task. 

The largest workflow change between Galaxy Zoo versions was between the original Galaxy Zoo (GZ1) and Galaxy Zoo 2 (GZ2). GZ1 presented classifiers with a single task per galaxy, a choice between smooth/elliptical, multiple versions of featured/disk (including edge-on, face-on, and directionality of spiral structure), and merger. GZ2 re-classified the brightest quarter of the GZ1 sample in much greater detail, including a branched, multi-task decision tree. Subsequent changes to the decision tree for different versions of Galaxy Zoo have been mostly iterative in nature, driven in part by the data itself and in part by experience-based reflection which revealed minor adjustments that could help classifiers provide more accurate information. As an example of the former, a new branch was added for GZ-Hubble and GZ-CANDELS to capture information on star-forming clumps in classifications of higher-redshift galaxies. As an example of the latter, the final 2 tasks of GZ2 have been adjusted over multiple versions to facilitate reliable identification of rare features. Such adjustments have generally been minimized to avoid complicating comparisons with previous campaigns.

The decision tree used for GZD-1 and GZD-2 has three modifications vs. the Galaxy Zoo 2 decision tree \citep{Willett2013}. The `Can't Tell' answer to `How many spiral arms are there?' was removed, the number of answers to `How prominent is the central bulge?' was reduced from four to three, and `Is the galaxy currently merging, or is there any sign of tidal debris?' was added as a standalone question.

\begin{figure}
    \centering
    \includegraphics[width=\columnwidth]{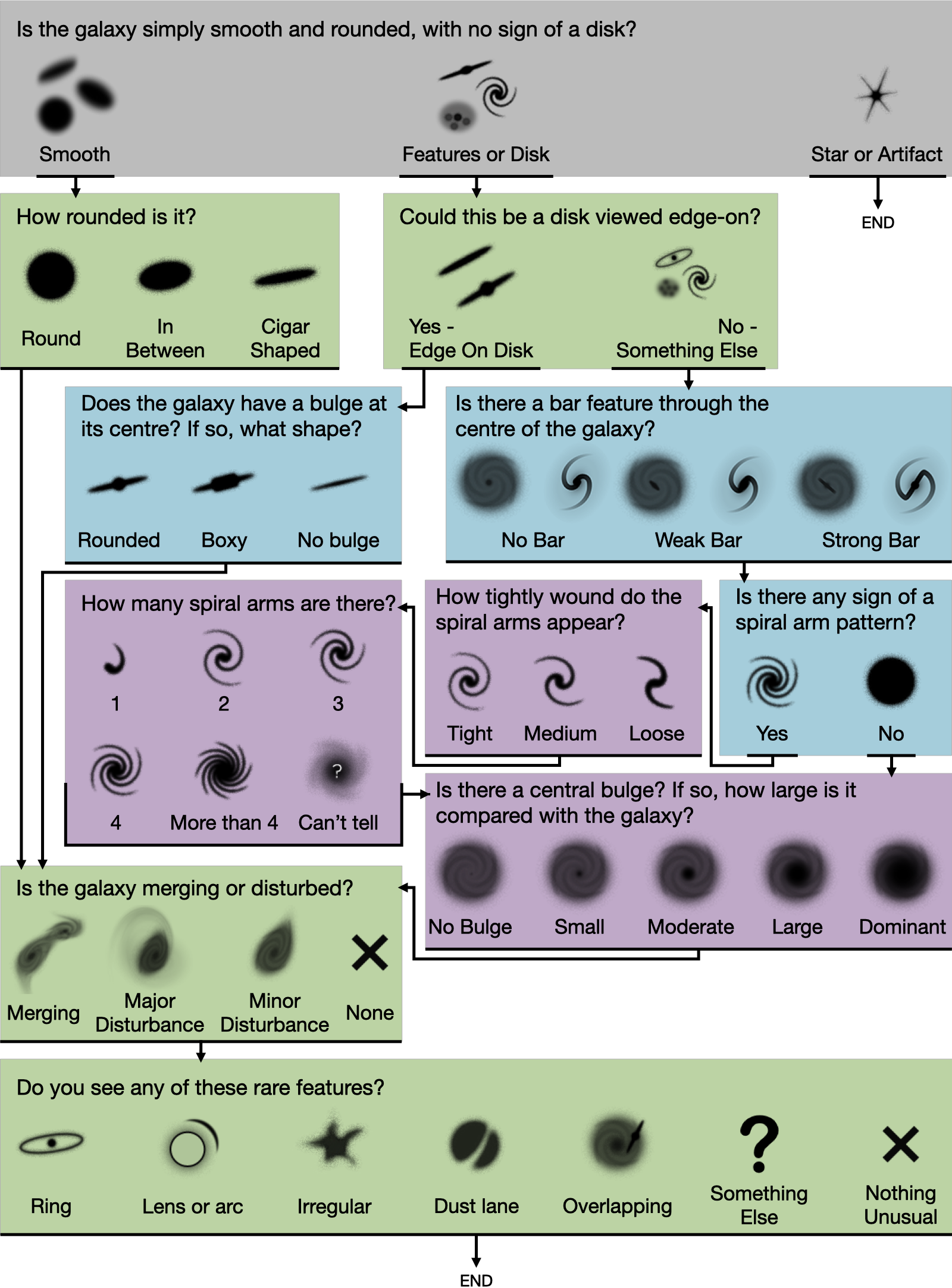}
    \caption{Classification decision tree for GZD-5, with new icons as shown to volunteers. Questions shaded with the same colours are at the same level of branching in the tree; grey have zero dependent questions, green one, blue two, and purple three.
    }
    \label{fig:decision_tree}
\end{figure}

For GZD-5, we made three further changes. Several Galaxy Zoo studies (e.g. \citealt{Skibba2012, Masters2012, Willett2013, Kruk2018}) found that galaxies selected with 0.2<$p_\mathrm{bar}$<0.5 in GZ2 correspond to `weak bars' when compared with expert classification such as those in \cite{Nair2010}. Therefore, to increase the detection of bars, we changed the possible answers to the `Does this galaxy have a bar?' question from `Yes' or `No' to `Strong', `Weak' or `No'. We define a strong bar as one that is clearly visible and extending across a large fraction of the size of the galaxy. A weak bar is smaller and fainter relative to the galaxy, and can appear more oval than the strong bar, while still being longer in one direction than the other. Our definition of strong vs. weak bar is similar that of \cite{Nair2010}, with the exception that they also have an `intermediate' classification. We added examples of galaxies with `weak bars' to the Field Guide and provided a new icon for this classification option, as shown in Figure \ref{fig:decision_tree}.

Second, to allow for more fine-grained measurements of bulge size, we increased the number of `How prominent is the central bulge?' answers from three (`No', `Obvious', `Dominant') to five (`No Bulge', `Small', `Moderate', `Large', `Dominant'). We also re-included the `Can't Tell' answer.

Third, we modified the `Merging' question from `Merging', `Tidal', `Both', or `None', to the more phenomenological `Merging', `Major Disturbance', `Minor Disturbance', or `No'. Our goal was to present more direct answers to our volunteers and to better distinguish major and minor mergers, to support recent scientific interest in the role of major and minor mergers on mass assembly \citep{Lopez-Sanjuan2010, Kaviraj2013Minor}, black hole accretion \citep{Alexander2012, Simmons2017a}, and morphology \citep{Hopkins2009a, Lotz2011, Lofthouse2017}. We made this final `merger' change two months after launching GZD-5; 6722 GZD-5 galaxies (2.7\%) were fully classified before that date and so do not have responses from volunteers to this question.

We also make several improvements to the illustrative icons shown for each answer. These icons are the most visible guide for volunteers as to what each answer means (complementing the tutorial, help text, field guide, and `Talk' forum). Figure \ref{fig:decision_tree} shows the GZD-5 decision tree with new icons as shown to volunteers. The decision tree used in GZD-1 and GZD-2 is shown in Figure \ref{fig:gzda_tree}.

For the `Smooth or Featured?' question, we changed the `Smooth' icon to include three example galaxies at various ellipticities, and the `Featured' icon to include an edge-on disk rather than a ring galaxy. For `Edge On?', we replaced the previous tick icon with a new descriptive icon, and the previous cross icon with the `Featured' icon above. We also modified the text to no longer specify `exactly' edge on, and renamed the answers from `Yes' and `No' to `Yes - Edge On Disk' and 'No - Something Else'. For `Bulge?', we created new icons to match the change from four to five answers. For `Bar', we replaced the previous tick and cross icons with new descriptive icons for `Strong Bar', `Weak Bar' and `No Bar'. For `Merger?', we added new descriptive icons to match the updated answers.

Changes to the decision tree complicate comparisons other Galaxy Zoo projects. As we show in the following sections, the available answers will affect the sensitivity of volunteers to certain morphological features, and so morphology measurements made with different decision trees may not be directly comparable. This difficulty in comparison has historically required us to be conservative in our changes to the decision tree. However, the advent of effective automated classifications allows us to retrospectively make classifications using any preferred decision tree. Specifically, in this work, we train our automated classifier to predict what volunteers would have said using the GZD-5 decision tree, for galaxies which were originally classified by volunteers using the GZD-1/2 decision tree (Section \ref{sec:bayesian_classifier}).

\section{Volunteer Analysis}
\label{sec:volunteer_analysis}

\subsection{Improved Feature Detection from DECaLS imagery}
\label{comparison_to_gz2}

The images used in GZ DECaLS are deeper and higher resolution than were available for GZ2. The GZ2 primary sample \citep{Willett2013} uses images from SDSS DR7 \citep{Abazajian2009}, which are 95\% complete to $r=22.2$ with a median seeing of 1 \farcs 4 and a plate scale of 0\farcs396 per pixel \citep{York2000}. In contrast, GZ DECaLS uses images from DECaLS DR2 to DR5, which have a median 5$\sigma$ point source depth of $r=23.6$, a seeing better than 1\farcs3 for at least one observation, and a plate scale of 0\farcs262 per pixel \citep{Dey2018}\footnote{See also http://www.legacysurvey.org/dr5/description/}.

We expect the improved imaging to reveal morphology not previously visible, particularly for features which are faint (e.g. tidal features, low surface brightness spiral arms) or intricate (e.g. weak bars, flocculent spiral arms). Our changes to the decision tree (Sec. \ref{sec:decision_trees_intro}) were partly made to better exploit this improved imaging.

To investigate the consequences of improved imaging, we compare galaxies classified in both GZ2 and GZ DECalS. Galaxies will typically be classified by both projects if they are inside both the SDSS DR7 Legacy catalogue (i.e. the source GZ2 catalogue) and DECaLS DR5 footprints (broadly, North Galactic Cap galaxies with $-35 < \delta < 0$) and match the selection criteria of each project (see \citealt{Willett2013} and Sec. \ref{sec:selection}). GZ2's $r < 17.0$ cut, with no corresponding GZ DECaLS magnitude cut, means that the the odds of any given GZ2 galaxy being in GZ DECaLS is close to random (for an isotropic sky) but only the brighter half of suitably-located GZ DECaLS galaxies are in GZ2. To exclude the effect of modifying the decision tree in GZD-5 (addressed separately in Sec \ref{sec:comparison_of_decision_trees}), we use only GZ DECaLS classifications from GZD-1 and GZD-2. 33,124 galaxies were classified in both GZ2 and GZD-1 or GZD-2.

\begin{figure}
    \centering
    \includegraphics[width=\columnwidth]{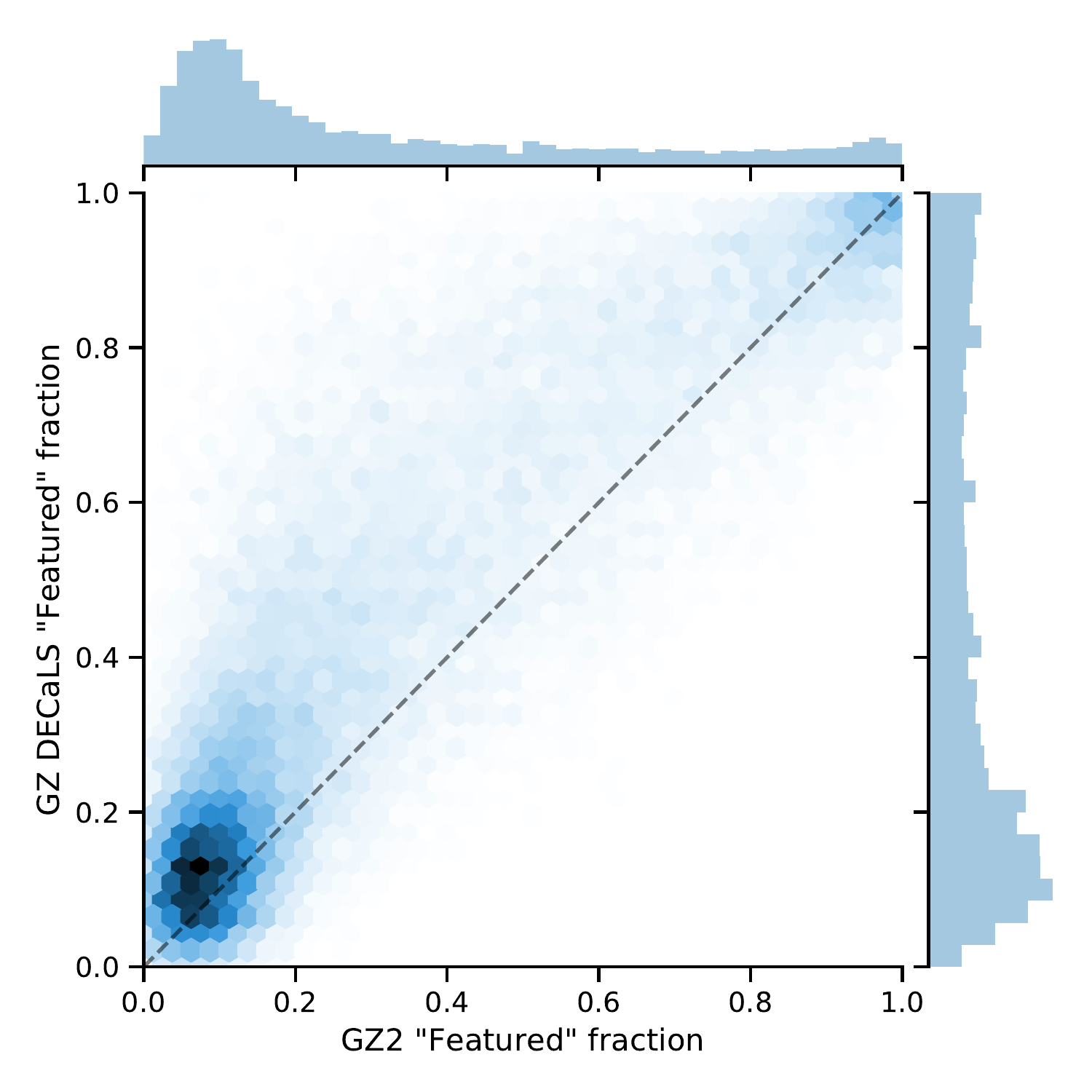}
    \caption{Comparison of `Featured' fraction for galaxies classified in both GZ2 and GZ DECaLS. Ambiguous galaxies are consistently reported as more featured in GZ DECaLS, which we attribute to the significantly improved imaging depth of DECaLS.}
    \label{fig:featured_comparison}
\end{figure}

We find that volunteers successfully recognise newly-visible morphology features. Figure \ref{fig:featured_comparison} compares the distribution of vote fractions to `Is this galaxy smooth or featured?' for GZ2 and GZ DECaLS. Ambiguous galaxies, with `featured' fractions (before debiasing) between approx. 0.25 and 0.75, are consistently reported as more featured (median absolute increase of 0.13, median percentage increase of 22\%) with the deeper GZ DECaLS images.

\begin{figure}
    \includegraphics[width=\columnwidth,trim={0 5cm 0 5cm},clip]{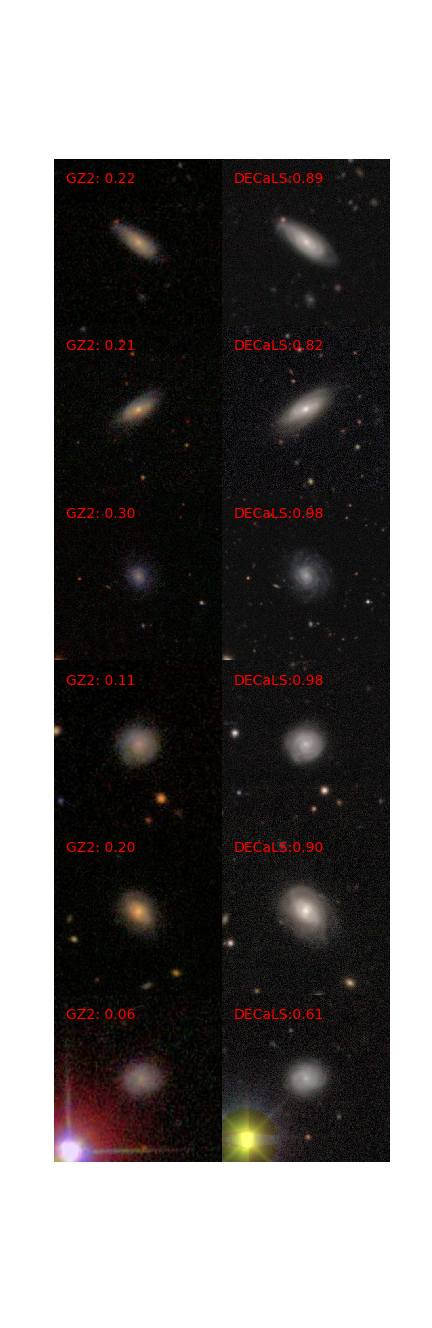}
    \caption{GZ2 and GZ DECaLS images for 6 galaxies drawn randomly from the 1000 galaxies classified in both projects with the largest increase in `featured' vote fraction (reported fractions shown in red). The increased fraction accurately reflects the increased visibility of detailed morphology from improved imaging.}
    \label{fig:featured_galaxies_big_shift}
\end{figure}

The shift towards featured galaxies is an accurate response to the new images, rather than systematics from (for example) a changing population of volunteers. Figure \ref{fig:featured_galaxies_big_shift} compares the GZ2 and GZ DECaLS images of a random sample of galaxies drawn from the 1000 cross-classified galaxies with the largest increase in `featured' fraction. In all of these galaxies (and for a clear majority of galaxies in similar samples), volunteers are correctly recognising newly visible detailed features.

We observe a similar pattern in the vote fractions of spiral arms and bars for featured galaxies. For galaxies consistently considered featured (i.e. where both projects reported a `featured' vote fraction of at least 0.5), the median vote fraction for spiral arms increased from 0.84 to 0.9, and for bars from 0.21 to 0.24. This suggests that even for galaxies where some details were already visible (and hence were considered featured), improved imaging makes our volunteers more likely to identify specific features.

We argue the improved depth of DECaLS ($r=23.6$ vs $r=22.2$ for SDSS) is revealing low surface brightness features that were previously ambiguous. There may also be contributions from the modified image processing approach and from the shift between using $gri$ bands (SDSS) to $grz$ bands (DECaLS), which might make older stars more prominent. 

Comparing classifications made using the same possible answers on the same galaxies shows how improved DECaLS imaging leads to ambiguous galaxies being correctly reported as more featured, and to spiral arms and bars being reported with more confidence. However, volunteers are also sensitive to which questions are asked and how those questions are asked. We measure the impact of our changes to the decision tree `Bar' question for GZD-5 in the next section.

\subsection{Improved Weak Bar Detection from GZD-5 Decision Tree}
\label{sec:comparison_of_decision_trees}

To measure the effect of the new decision tree on bar sensitivity, we compare the classifications made using each tree against expert classifications. \citealt{Nair2010} (hereafter NA10) classified all 14,034 SDSS DR4 galaxies at $0.01 < z < 0.05$ with $g < 16$. Of those, 1497 were imaged by DECaLS DR1/2 and classified by volunteers during GZD-1/2. We re-classified these galaxies during GZD-5 to measure the effect of the new bar answers, as compared to the expert classifications of NA10. Note that because NA10 used shallower SDSS images, NA10's classifications are best used as positive evidence; while NA10 finding a bar in SDSS images implies a visible bar in DECaLS images, NA10 not finding a bar may not always exclude a visible bar in DECaLS. To exclude smooth galaxies, which are unbarred by definition in our schema, we require $f_{\text{featured}} > 0.25$ (as measured by GZD-5), selecting a featured sample of 807 galaxies classified by NA10, GZD-1/2, and GZD-5. 

\begin{figure}
    \centering
    \includegraphics[width=\columnwidth]{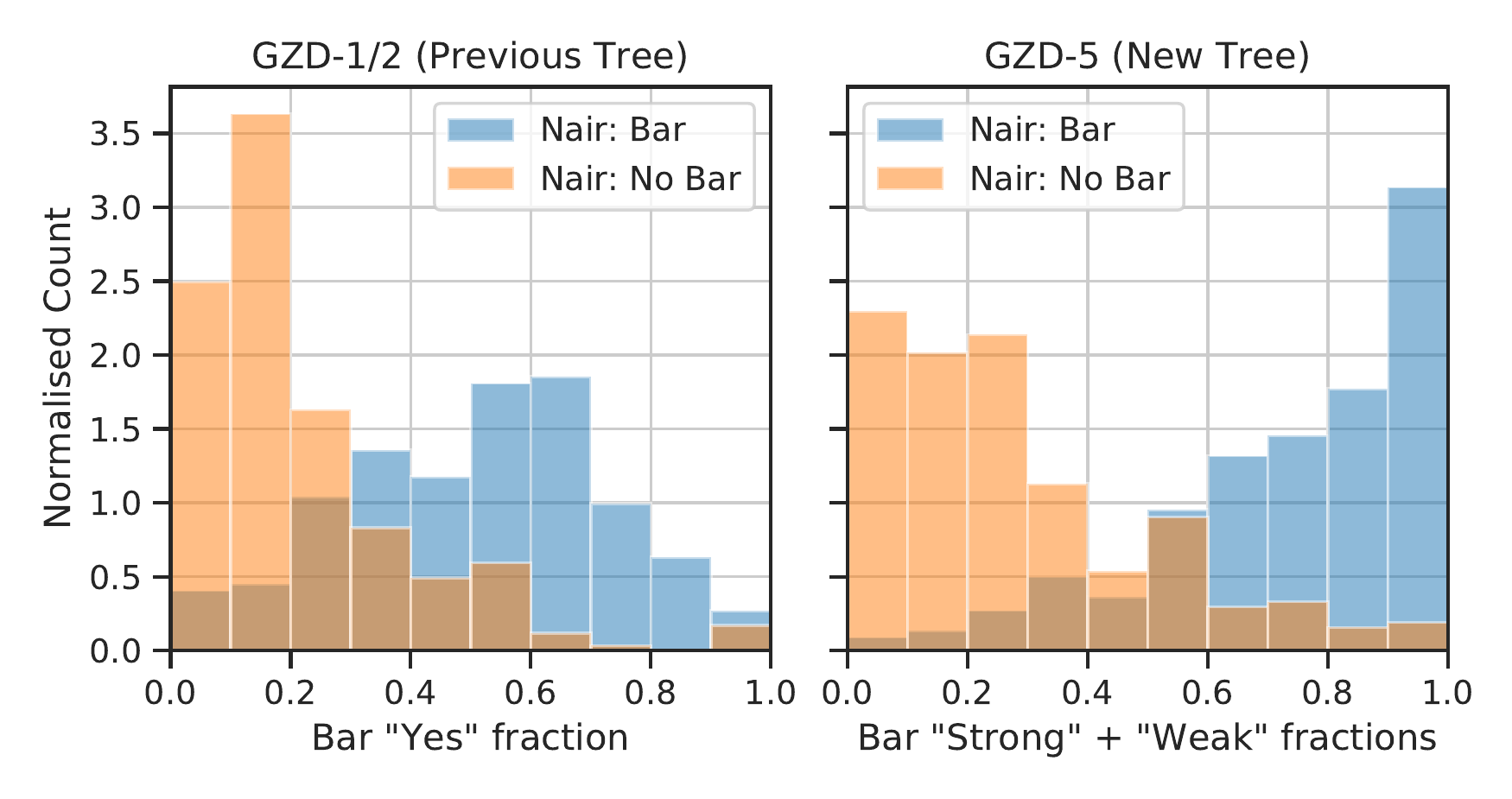}
    \caption{
    Left: Distribution of fraction of GZD-1/2 volunteers answering `Yes' (not `No' to `Does this galaxy have a bar?', split by expert classification from NA10 of barred (blue) or unbarred (orange). Right: as left, but for GZD-5 volunteers answering `Strong' or `Weak' (not `No'). Volunteers are substantially better at identifying barred galaxies using the GZD-5 three-answer question.
    }
    \label{fig:bar_any_comparison}
\end{figure}

Figure \ref{fig:bar_any_comparison} compares volunteer classifications for expert-labelled calibration galaxies made using each tree. We find that barred and unbarred galaxies are significantly better separated with the Strong/Weak/None answers than with Yes/No answers. Of 220 Nair-identified bars (of any type), 184 (84\%) receive a majority vote for being barred by volunteers using the new tree, up from 120 (55\%) with the previous tree.

\begin{figure}
    \centering
    \includegraphics[width=\columnwidth]{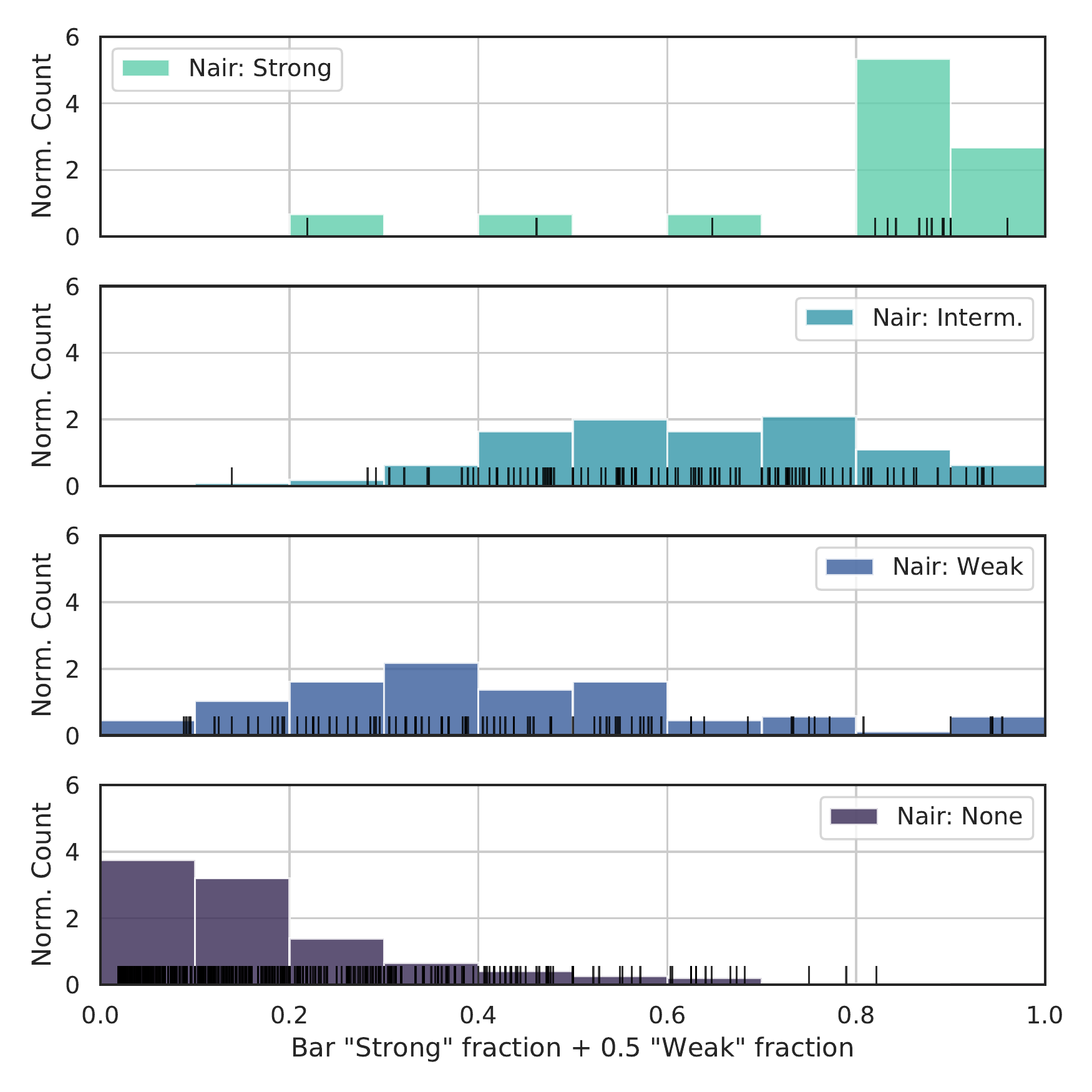}
    \caption{
        Distributions of volunteer bar strength estimates,  $B_\text{vol} = f_{\text{strong}} + 0.5 f_{\text{weak}}$, split by expert-classified (NA10) bar strength. Individual galaxies are shown with rug plots (15 Strong, 110 Intermediate, 87 Weak, and 377 None). Volunteer bar strength estimates increase smoothly with expert-classified bar strength, though individual galaxies vary substantially. 
    }
    \label{fig:bar_answer_comparison}
\end{figure}

NA10 classified barred galaxies into five subtypes: Strong, Intermediate, Weak, Nuclear, Ansae, and Peanut (plus None, implicitly). We can use the first three subtypes as a measurement of expert-classified bar strength, and therefore evaluate how our volunteers respond to bars of different strengths. Following the approach to defining summary metrics of \cite{Masters2019}, we summarise the bar vote fractions into a single volunteer estimate of bar strength,  $B_\text{vol} = f_{\text{strong}} + 0.5 f_{\text{weak}}$, and compare the distribution of $B$ for each expert-classified bar strength (Figure \ref{fig:bar_answer_comparison}). We find that the volunteer bar strength estimates increase smoothly with expert-classified bar strength, though individual galaxies vary substantially. This suggests that typical bar strength in galaxy samples can be successfully inferred from volunteer votes. 

The addition of the `weak bar' answer in GZD-5 significantly improves sensitivity to bars compared with previous versions of the decision tree. Additionally, volunteer votes across the three answers may be used to infer bar strength. We hope that the detailed bar classifications in our catalogue will help researchers better understand the properties of strong and weak bars and their influence on host galaxies.

\subsection{Classification Modifications}
\label{sec:classification_modificatons}

Galaxy Zoo data releases have previously included two post-hoc modifications to the volunteer classifications; volunteer weighting, to reduce the influence of strongly atypical volunteers, and redshift debiasing, to estimate the vote fractions a galaxy might have received had it been observed at a specific redshift. We describe each modification below.

\subsubsection{Volunteer Weighting}
\label{sec:volunteer_weighting}

Volunteer weighting, as introduced in Galaxy Zoo 2 \citep{Willett2013}, assigns each volunteer an aggregation weight of (initially) one, and iteratively reduces that weight for volunteers who typically disagree with the consensus. This method affects relatively few volunteers and therefore causes only a small shift in vote fractions - in Galaxy Zoo 2, for example, approximately 95\% of volunteers had a weighting of one (i.e. unaffected), 94.8\% of galaxies had a change in vote fraction of no more than 0.1 for any question, and the mean change in vote fraction across all questions and galaxies was 0.0032. 

The most significant change in final vote fractions is caused by down-weighting rare (approx. 1\%) volunteers who repeatedly disagree with consensus by answering `artifact' at implausibly high rates (including 100\%) for many galaxies. Answering artifact ends the classification and shows the next galaxy, and so we hypothesise that these rare volunteers are primarily interested in seeing many galaxies rather than contributing meaningful classifications. There are very few such volunteers, but because answering artifact allows classifications to be submitted very quickly, they have an outsize effect on the aggregate vote fractions. 

\begin{figure}
    \centering
    \includegraphics[width=\columnwidth]{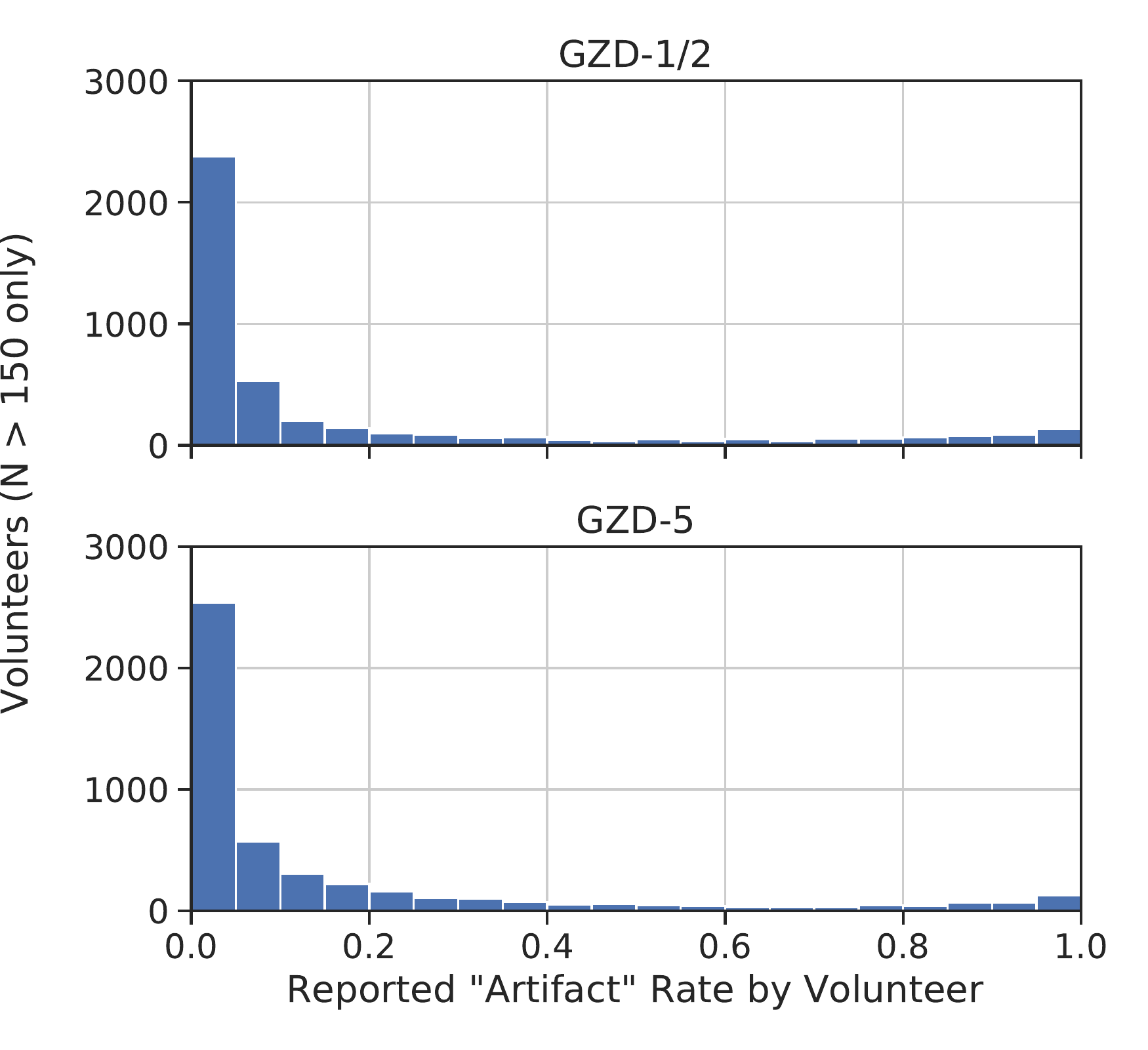}
    \caption{
        Distribution of reported `artifact' rates by volunteer (i.e. how often each volunteer answered `artifact' over all the galaxies they classified). The vast majority report artifact rates consistent with those of the authors (below 0.1), but a very small subset report implausibly high artifact rates $(> 0.5)$ and consequently have their classifications discarded. Only volunteers with at least 150 classifications are shown; the distribution for volunteers with fewer classifications is not bimodal.
    }
    \label{fig:artifact_fractions}
\end{figure}

Figure \ref{fig:artifact_fractions} shows the distribution of reported artifact rates for volunteers with at least 150 total classifications. We expect the true fraction of artifacts to be less than 0.1, and the vast majority of volunteers report artifact rates consistent with this. However, the distribution is bimodal, with a small second peak around 1.0 (i.e. volunteers reporting every galaxy as an artifact). To remove the implausible mode, we discard the classifications of volunteers with at least 150 total classifications and reported artifact rates greater than 0.5. In GZD-1/2, 1.1\% (643) of volunteers are excluded, discarding 11\% (483,081) of classifications. In GZD-5, 0.03\% (543) volunteers are excluded, discarding 5.3\% (249,592) of classifications. 

We investigated the possibility of other groups of atypical volunteers giving similar answers across questions by analysing the per-user vote fractions with either a two-dimensional visualisation using UMAP \citep{McInnes2018} or with clustering using HDBSCAN \citep{McInnes2017}. We find no strong evidence that such clusters exist.

\subsubsection{Redshift Debiasing}
\label{sec:redshift_debiasing}

\begin{figure}
    \centering
    \includegraphics[width=\columnwidth]{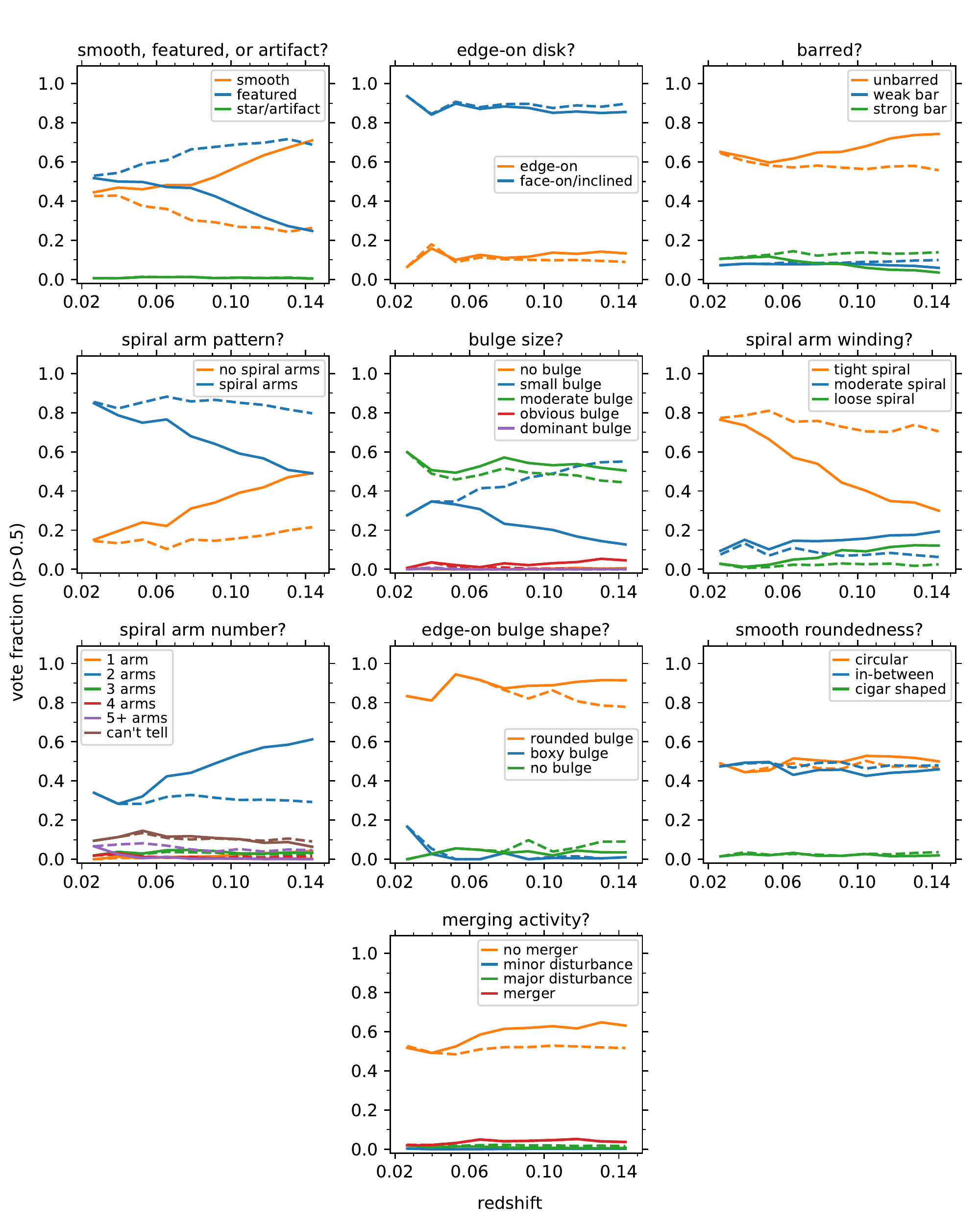}
    \caption{Number of GZD-5 galaxies with $f > 0.5$ for each of the questions debiased using the method described \ref{sec:redshift_debiasing}. The solid lines indicate the original vote fractions and the dashed lines indicate the debiased vote fractions. The total sample here is composed of galaxies in the luminosity-limited sample with $f > 0.5$, 58,916 galaxies. For most questions and answers, debiasing successfully flattens the redshift trends. For `Smooth or Featured' and `Bulge Prominence', redshift debiasing overcorrects.}
    \label{fig:debiasing}
\end{figure}

Galaxies at higher redshifts appear fainter and smaller on the sky, making it harder to detect detailed morphological features than if the galaxy were closer. This creates a bias in visual classifications (whether human or automated) where galaxies of the same intrinsic morphology are less likely to be classified as having detailed features as redshift increases \citep{Bamford2009}. Redshift debiasing is an attempt to mitigate this bias by estimating how a galaxy would appear if it were at a fixed low redshift (here, $z=0.02$). 

We use the method described in \citet{Hart2016} to remove the redshift bias, which we briefly summarise here and refer the reader to their Section 3 for full details. We assume the morphological properties of galaxies (as probed by our decision tree) over the redshift window covered by Galaxy Zoo DECaLS ($0.02<z<0.15$, approximately 1.5 Gyr) do not evolve significantly for galaxies of similar intrinsic brightness and physical size, and so, for a luminosity-limited sample, any change we observe to the vote fraction distribution as a function of redshift is purely a consequence of imaging. If so, we can estimate the vote fractions which would be observed if each galaxy were at low redshift by modifying the vote fractions of higher-redshift galaxies such that they have the same overall distribution as their low-redshift counterparts in brightness and size.

We base the debiasing on a \textit{luminosity-limited sample}, selected between $0.02<z<0.15$ and $-21.5>M_r>-23$. We consider the galaxies with at least 30 votes for the first question (`Smooth or Featured') after volunteer weighting (above), for a total of 87,617 galaxies in GZD-1/2 and 58,916 galaxies in GZD-5. For each question, separately, we define a subset of galaxies to which we apply the debiasing procedure. 

Each subset is defined using a cut of $f > 0.5$ for the chain of preceding questions (for example, for the bar question, we require $f_{\text{feat}} \times f_{\text{not edge-on}} > 0.5$). A further cut of $N > 5$ (where N is the number of classifications) is also imposed to ensure that each galaxy has been classified by a significant number of people. We bin the subset of galaxies by $M_r$, $\log(R_{50})$ and $z$ for each answer in turn. We use the \texttt{voronoi\_2d\_binning} package from \cite{Cappellari2003} to ensure the bins will have an approximately equal number of galaxies (with a minimum of 50). We then match vote fraction distributions on a bin-by-bin basis, such that the cumulative distribution of vote fractions at each redshift is shifted to be similar to that of the lowest redshift sample ($0.02<z<0.03$). This method aims to keep the fraction of galaxies above a given threshold constant with redshift. 

The effect of redshift bias and redshift debiasing question is shown in Fig. \ref{fig:debiasing}. To illustrate, consider the `Smooth or Featured' question (top left). In a luminosity-limited sample, there should be the same fraction of galaxies with features (selected with $f_{\text{feat}} > 0.5$) at all redshifts. However, we observe that the fraction of `featured' galaxies decreases, and the fraction of `smooth' galaxies increases (solid lines). We attribute this to redshift bias; some galaxies that would be considered featured if imaged at low redshift appear as `fuzzy blobs' at high redshift and are instead classified as smooth. After applying redshift debiasing, the debiased fractions (dashed lines) change more gradually with redshift. For most questions and answers, the redshift trend is successfully flattened (recall that for every size and luminosity bin, we enforce no change in the vote fraction distribution with redshift). For `Smooth or Featured' and `Bulge Prominence', the debiasing procedure overcorrects and hence reverses the redshift trend.

For statistical studies it is important to test for the presence of a classification bias with redshift and correct it where necessary. Such a correction has proven essential in studies of the morphology density relation \citep{Bamford2009} and when characterising populations with different spiral arm properties \citep{Hart2016}. However, while debiasing can be extremely useful, there are caveats to its usage. It is sometimes helpful to think of the original classifications as a lower limit to the probability of features of a given type existing in a galaxy. Debiasing predicts what the classifications would be if the same galaxy were imaged at lower redshift, which is typically more featured than the original classifications. There is substantial uncertainty in this prediction, however, and this is currently not captured by the debiased vote fractions, which are reported without error bars.

In some investigations it may be helpful to consider that the true classification for a given galaxy is likely to be in between the original classification and the debiased classification. At the same time, the debiased classifications are not strictly upper limits. They are based on the lowest-redshift classifications within the dataset itself, which themselves are at a non-zero redshift, and so there are likely differences in the debiased classifications and the `true' debiased classification that would be assigned if we could image the galaxy at arbitrarily-low redshift. As these corrections are applied uniformly, however, they are useful when considering overall populations of galaxies within a given dataset and over the redshift ranges where the correction is relevant. In particular, when \emph{comparing} different morphological types, some of the systematic errors in the debiasing may cancel out. Uncertainties in the debiasing will also decrease as the sample size increases.

For these reasons, we strongly suggest that users of the debiased classifications only use them to consider populations of galaxies rather than individual or small samples, and to consider that there may still be some residual trends and uncertainties that are hard to model with current methods. 

\section{Automated Classifications}
\label{sec:automated}

Combining citizen science with automated classification allows us to do better science than with either alone.
The clearest benefit is that automated classification scales well with sample size. For GZ DECaLS, classifying all 311,488 suitable galaxies using volunteers alone is infeasible; collecting 40 classifications per galaxy, the standard from previous Galaxy Zoo projects, would take around eight years without further promotion efforts - by which time we expect new surveys to start. Automated classification also evolves - as the quality of our models improves, so too will the quality of our classifications. And automated classification is replicable from scratch without requiring a crowd - other researchers may run our open-source code and recover our classifications (within stochasticity), or create equivalent classifications for newly-imaged galaxies. 

Finally, and of particular relevance to researchers using this data release, automated classification allows us to retroactively update the decision tree. Because our classifier learns to make predictions from GZD-5 classifications, using the improved tree with better detection of mergers and weak bars, we can then predict what our volunteers would have said for the GZD-1 and GZD-2 galaxies \textit{had we been using the improved tree at that time}.

Our specific automated classification approach offers several qualitative benefits over previous work. First, through careful consideration of uncertainty, we can both learn from uncertain volunteer responses and predict posteriors (rather than point estimates) for new galaxies.  Second, by predicting the answers to every question with a single model (similarly to \citealt{Dieleman2015}, and unlike more recent work e.g. \citealt{Sanchez2018, Khan2018, Walmsley2020}), we improve performance by sharing representations between tasks \citep{Caruana1997} - intuitively, knowing how to recognise spiral arms can also help you count them. Learning from every galaxy to predict every answer uses our valuable volunteer effort as efficiently as possible. This is particularly effective because we aim to predict detailed morphology, and hence learn to create a detailed representation of each galaxy.

\subsection{Bayesian Deep Learning Classifier}
\label{sec:bayesian_classifier}

We require a model which can:
\begin{enumerate}
    \item Learn efficiently from volunteer responses of varying (i.e. heteroskedastic) uncertainty.
    \item Predict posteriors for those responses on new galaxies, for every question.
\end{enumerate}

In previous work \citep{Walmsley2020} we modelled volunteer responses as being binomially distributed and trained our model to make maximum likelihood estimates using the loss function

\begin{equation}
    \mathcal{L} = k \log f^w(x) + (N-k) \log(1-f^w(x))
\end{equation}
where, for some target question, $k$ is the number of responses (successes) of some target answer, $N$ is the total number of responses (trials) to all answers, and $f^w(x) = \hat{\rho}$ is the predicted probability of a volunteer giving that answer.

This Binomial assumption, while broadly successful, broke down for galaxies with vote fractions $\frac{k}{N}$ close to 0 or 1, where the Binomial likelihood is extremely sensitive to $f^w(x)$, and for galaxies where the question asked was not appropriate (e.g. predict if a featureless galaxy has a bar). Instead, in this work, the model predicts a distribution $p(\rho|f^w(x))$ and $\rho$ is then drawn from that distribution.

For binary questions, one could parametrise $p(\rho|f^w(x))$ with the Beta distribution (being flexible and defined on the unit interval), and predict the Beta distribution parameters $f^w(x) = (\hat{\alpha}, \hat{\beta})$ by minimising

\begin{equation}
    \mathcal{L} = \int \text{Bin}(k|\rho, N) \text{Beta}(\rho|\alpha, \beta) d\alpha d\beta
\end{equation}
where the Binomial and Beta distributions are conjugate and hence this integral can be evaluated analytically.

In practice, we would like to predict the responses to questions with more than two answers, and hence we replace each distribution with its multivariate counterpart; Beta($\rho|\alpha, \beta$) with Dirichlet($\vec{\rho}|\vec{\alpha})$, and Binomial($k|\rho, N$) with Multinomial($\vec{k}|\vec{\rho}, N$).

\begin{equation}
    \label{multivariate_per_q_likelihood}
    \mathcal{L}_q = \int \text{Multi}(\vec{k}|\vec{\rho}, N) \text{Dirichlet}(\vec{\rho}| \vec{\alpha}) d\vec{\alpha}
\end{equation}
where $\vec{k}, \vec{\rho}$ and $\vec{\alpha}$ are now all vectors with one element per answer. 

The Dirichlet-Multinomial distribution is much more flexible than the Binomial, allowing our model to express uncertainty through wider posteriors and confidence through narrower posteriors. We believe this is a novel approach. 

For the base architecture, we use the EfficientNet B0 model \citep{Tan2019a}. The EfficientNet family of models includes several architectural advances over the standard convolutional neural network architectures commonly used within astrophysics (e.g. \citealt{Huertas-Company2015a, Dieleman2015, Khan2018, Cheng2019, Ferreira2020}), including auto-ML-derived structure \citep{Tan2018, He2019}, depthwise convolutions \citep{Howard2017}, bottleneck layers \citep{Iandola2016}, and squeeze-and-excitation optimisation \citep{Hu2018}. The EfficientNet B0 model was identified using multi-objective neural architecture search \citep{Tan2018}, optimising for both accuracy and FLOPS (i.e. computational cost of prediction). This balancing of accuracy and FLOPS is particularly useful for astrophysics researchers with limited access to GPU resources, leading to a model capable of making reliable predictions on hundreds of millions of galaxies. In short, the architecture is similar to traditional convolutional neural networks, being composed of a series of convolutional blocks of decreasing resolution and increasing channels. Each convolutional block uses mobile inverted bottleneck convolutions following MobileNetV2 \citep{Sandler2018}, which combine computationally efficient depthwise convolutions with residual connections between bottlenecks (as opposed to residual connections between blocks with many channels, as in e.g. ResNet \citep{He2016}). EfficientNet B0 has 5.3 million parameters.

We modify the final EfficientNet B0 layer output units to give predictions smoothly between 1 and 100 (using softmax activation), which is appropriate for Dirichlet parameters $\vec{\alpha}$. $\vec{\alpha}$ elements below 1 can lead to bimodal `horseshoe' posteriors, and $\vec{\alpha}$ elements above approximately 100 can lead to extremely confident predictions in extreme $\rho$, both of which are implausible for galaxy morphology posteriors. These constraints may cause the most extreme galaxies to have predicted vote fractions which are slightly less extreme than volunteers would record, but we do not anticipate this to affect practical use; whether a galaxy is extremely likely to have a bar or merely highly likely is rarely of scientific consequence.

We would like our single model to predict the answer to every question in the Galaxy Zoo tree. To do this, our architecture uses one output unit per answer (i.e. for 13 questions with a total of 20 answers, we use 20 output units). We calculate the (negative log) likelihood per question (Eqn. \ref{multivariate_per_q_likelihood}), and then, treating the errors in the model's answers to each question as independent events, calculate the total loss as

\begin{equation}
    \log \mathcal{L} = \sum_q \mathcal{L}_q(\vec{k_q}, N_q, \vec{f^w_q})
\end{equation}
where, for question $q$, $N_q$ is the total answers, $\vec{k_q}$ is the observed votes for each answer, and $\vec{f^w_q}$ is the values of the output units corresponding to those answers (which we interpret as the Dirichlet $\vec{\alpha}$ parameters in Eqn. \ref{multivariate_per_q_likelihood}).

We train our model using the GZD-5 volunteer classifications. Because the training set includes both active-learning-selected galaxies receiving at least 40 classifications and the remaining GZD-5 galaxies with around 5 classifications, it is crucial that the model is able to learn efficiently from labels of varying uncertainty. Unlike \cite{Walmsley2020}, which trained one model per question and needed to filter galaxies where that question asked may not be appropriate, we can predict answers to all questions and learn from all labelled galaxies.

We train or evaluate our models using the 249,581 (98.5\%) GZD-5 galaxies with at least three volunteer classifications. Learning from galaxies with even fewer (one or two) classifications should be possible in principle, but we do not attempt it here as we do not expect galaxies with so few classifications to be significantly informative. The Dirichlet concentrations (distribution parameters) used to calculate our metrics are predicted by three identically-trained models, each making 5 forward passes with random dropout configurations and augmentations. We ensemble all 15 forward passes by simply taking the mean posterior given the total votes recorded, which may be interpreted as the posterior of an equally-weighted mixture of Dirichlet-Multinomial distributions. This mean posterior can then be used to calculate credible intervals (error bars) and in standard statistical analyses.  We develop our approach using a conventional 80/20 train-test split, and make a new split before calculating the final metrics reported here. 

For the published automated classifications, where we aim simply to make the best predictions possible rather than to test performance, we train on all 249,581 galaxies with at least 3 votes (98.5\%). We also train five rather then three models to maximise performance. Training each model on an NVIDIA V100 GPU takes around 24 hours. We then make predictions (using the updated GZD-5 schema) on all 313,789 galaxies in all campaigns. Each prediction (forward pass) takes approx. 6ms, equating to approx. 160ms for each published posterior.

Starting from the galaxy images shown to volunteers (Section \ref{sec:image_construction}), we take an average over channels to remove color information and avoid biasing our morphology predictions \citep{Walmsley2020}, then resize and save the images as 300x300x1 matrices. We then apply random augmentations when loading each image into memory, creating a unique randomly-modified image to be used as input to the network. We first apply random horizontal and vertical flips, followed by an aliased rotation by a random angle in the range (0, $\pi$), with missing pixels being filled by reflection on the boundaries. Finally, we  crop the image about a random centroid to 224x224 pixels, effectively zooming in slightly towards a random off-center point. We also apply these augmentations at test time to marginalise our posteriors over any unlearned variance. We train using the Adam \citep{Kingma2015} optimizer and a batch size of 128. We end training once the model loss fails to improve for 10 consecutive epochs.

Code for our deep learning classifier, including extensive documentation and several worked examples, is available at \href{https://github.com/mwalmsley/zoobot}{https://github.com/mwalmsley/zoobot}.

\subsection{Results}
\label{sec:automated_results}

\begin{figure}
    \centering
    \includegraphics[width=\columnwidth]{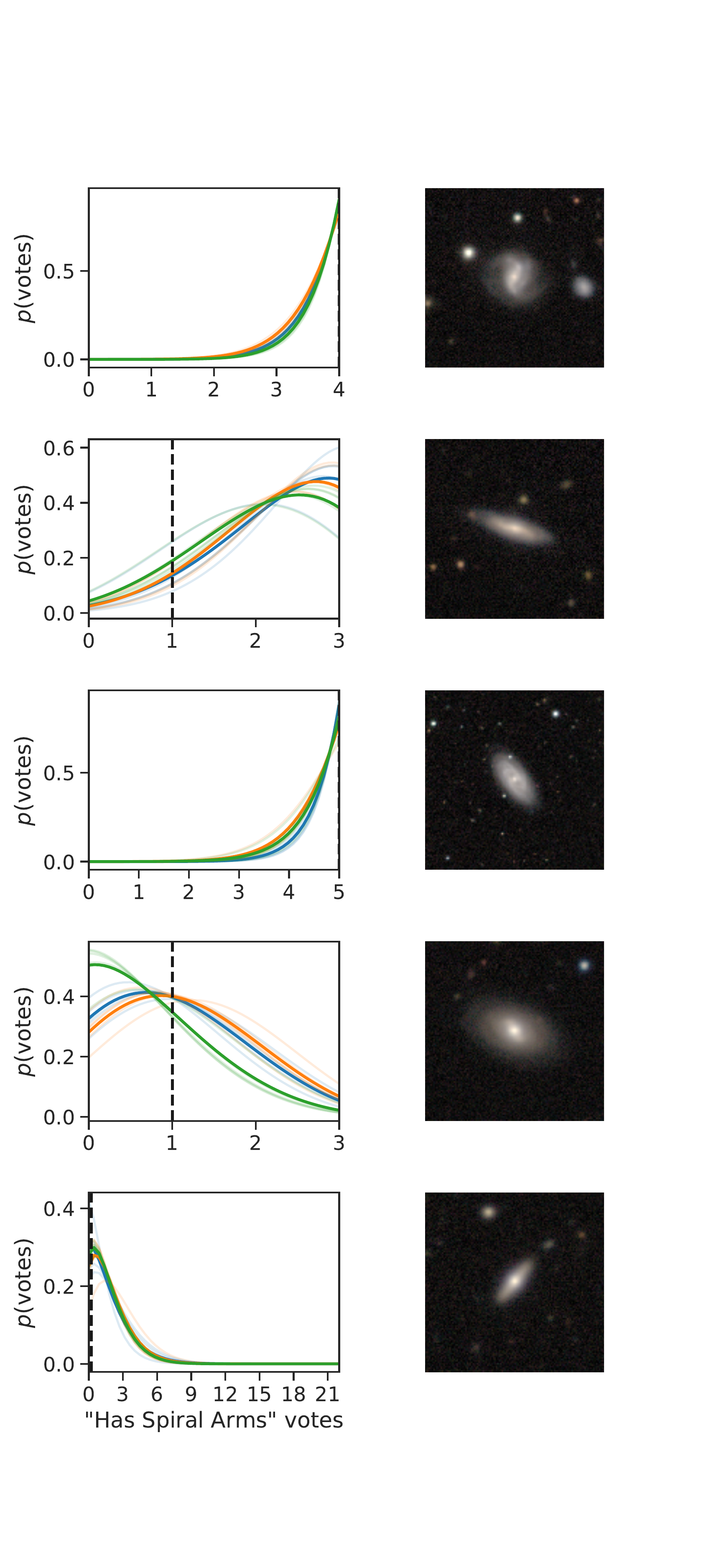}
    \caption{Posteriors for `Does this galaxy have spiral arms?', split by ensemble model (bold colours) and, within each model, dropout forward passes (faded colours). The number of volunteers answering `Yes' (not known to classifier) is shown with a black dashed line. Galaxies are selected at random from the test set, provided the spiral question is relevant (defined as a vote fraction of 0.5 or more to the preceding answer, `Featured').
    The image presented to volunteers is shown to the right. The model input is a cropped, downsized, greyscale version (Sec \ref{sec:bayesian_classifier}). The Dirichlet-Multinomial posteriors are strictly only defined at integer votes; for visualisation only, we show the $\Gamma$-generalised posteriors between integer votes.}
    \label{fig:example_posterior_spiral}
\end{figure}

Our model successfully predicts posteriors for volunteer votes to each question. We show example posteriors for a question with two answers, `Does this galaxy have spiral arms' (Yes/No), in Fig. \ref{fig:example_posterior_spiral}, and a question with three answers, `Does this galaxy have a bar' (Strong/Weak/None), in Fig. \ref{fig:example_posterior_bar}. In Appendix A, we provide a gallery of the galaxies with the highest expected vote fractions for a selection of answers, to visually demonstrate the quality of the most confident machine classifications.

\begin{figure}
    \centering
    \includegraphics[width=\columnwidth]{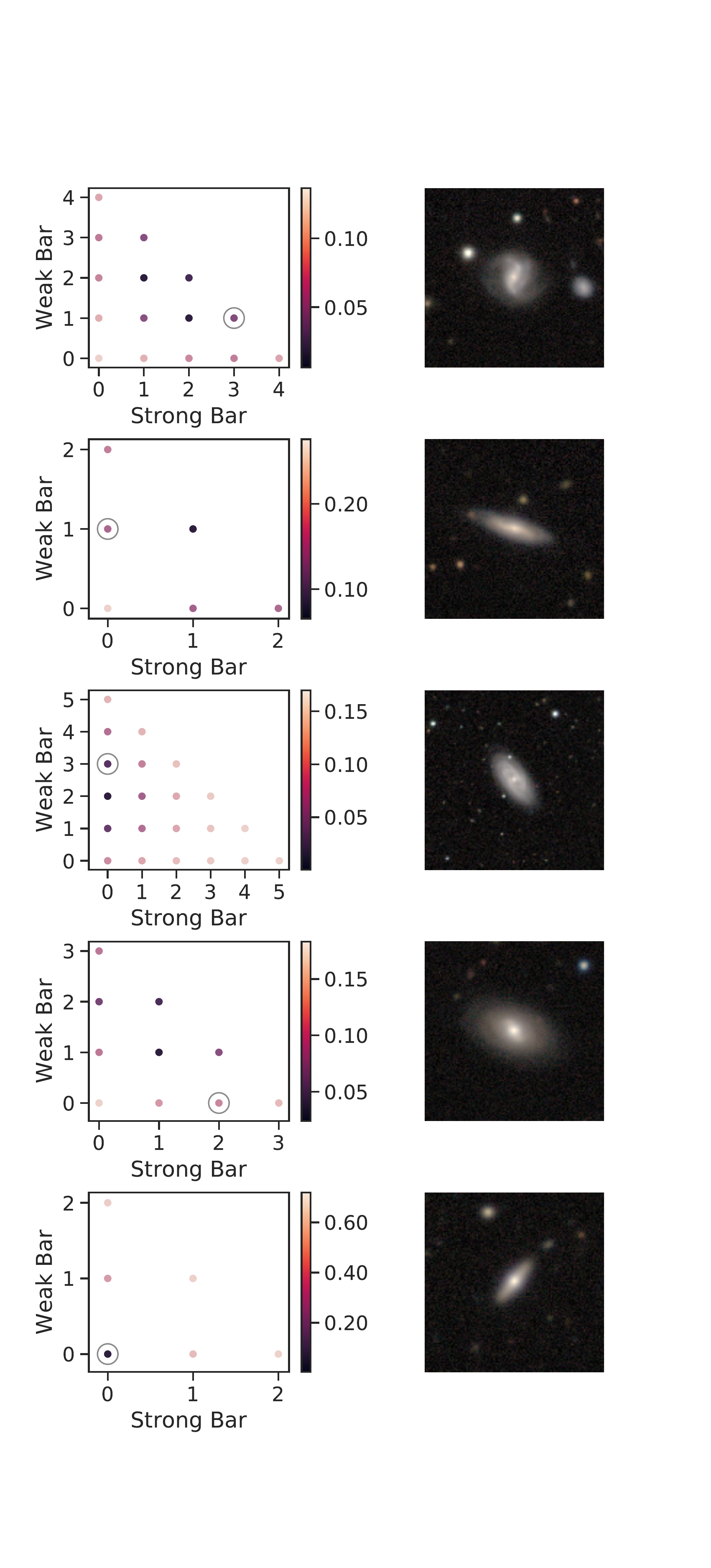}
    \caption{Posteriors for `Does this galaxy have a bar?', for the same random galaxies selected in Fig. \ref{fig:example_posterior_spiral}. Each point is colored by the predicted probability of volunteers giving that many `Strong', `Weak', and (implicitly, as the total answers is fixed) `None' votes. The volunteer answer (not known to classifier) is circled. For clarity, only the mean posterior across all models and dropout forward passes is shown.}
    \label{fig:example_posterior_bar}
\end{figure}

\begin{figure*}
    
    \includegraphics[width=.5\textwidth]{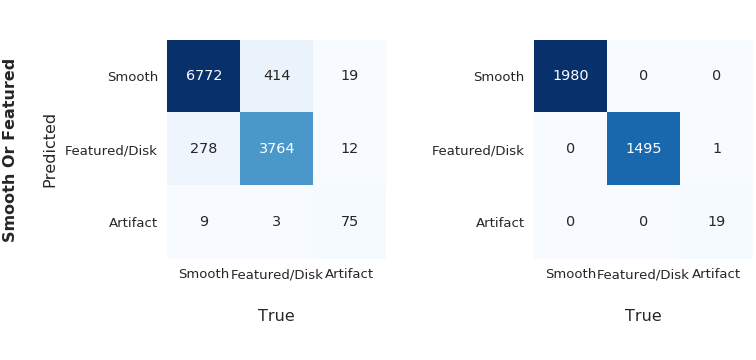}
        
    \includegraphics[width=.5\textwidth]{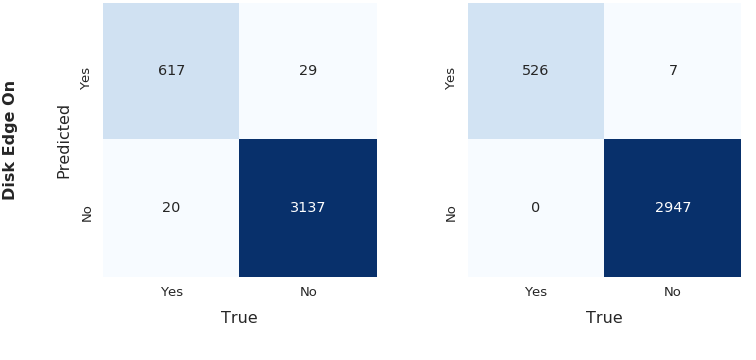}
    
    \includegraphics[width=.5\textwidth]{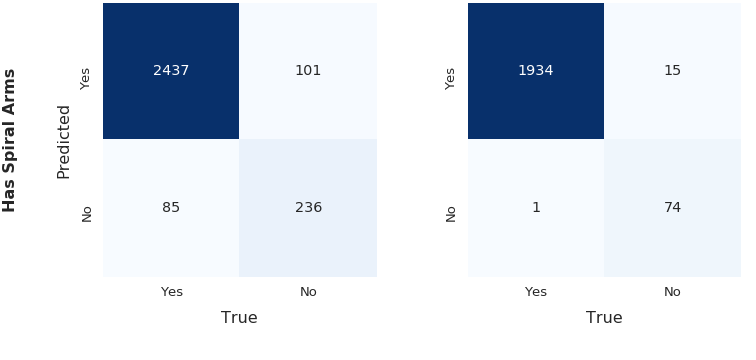}
    
    \includegraphics[width=.5\textwidth]{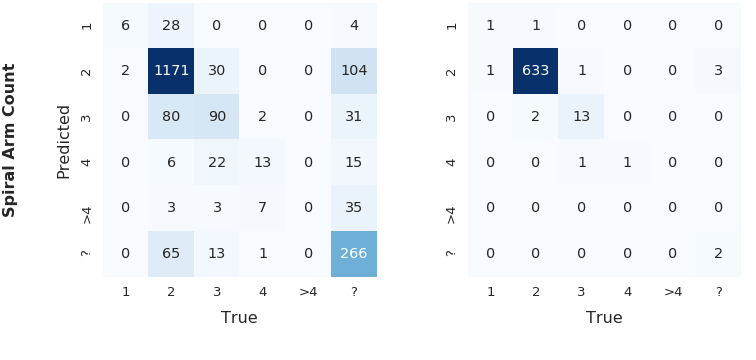}
    
    \includegraphics[width=.5\textwidth]{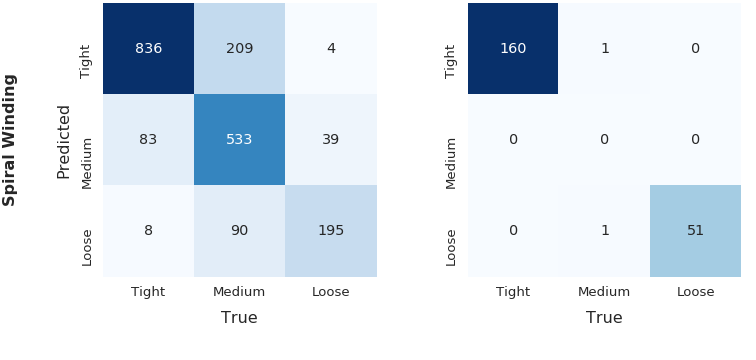}
    
    \hspace{52pt}\textit{All Galaxies} \hspace{52pt} \textit{High Volunteer Confidence}
    
    \caption{
        Confusion matrices for each question, made on the test set of 11,346 galaxies in the (random) test set with at least 34 votes. Discrete classifications are made by rounding the vote fraction (label) and mean posterior (prediction) to the nearest integer. The matrices then show the counts of rounded predictions (x axis) against rounded labels (y axis). To avoid the loss of information from rounding, we encourage researchers not to treat Galaxy Zoo classifications as discrete, and instead to use the full vote fractions or posteriors where possible.
        }
    \label{fig:confusion_matrices}
  \end{figure*}

To aid intuition for the typical performance, we reduce both the vote fraction labels and the posteriors down to discrete classifications by rounding the vote fractions and mean posteriors to 0 or 1, and calculate classification metrics (Table \ref{tab:ml_metrics}) and confusion matrices (Figure \ref{fig:confusion_matrices}). Here and throughout this section, we calculate performance on the 11,346 galaxies in the (random) test set with at least 34\footnote{Corresponding to the typical `full' retirement limit of approximately 40 classifications before discarding implausible classifications, see Sec. \ref{sec:volunteer_weighting}} classifications (such that the typical volunteer answer is well-measured). To remove galaxies for which the question is not relevant, we only count galaxies where at least half the volunteers were asked that question. We report two sets of classification metrics; metrics for all (relevant) galaxies, and only for galaxies where the volunteers are confident (defined as having a vote fraction for one answer above 0.8, following \citealt{DominguezSanchez2019}). 

The performance on confident galaxies is useful to measure because such galaxies have a clear correct label. For such galaxies, performance is near-perfect; we achieve better than 99\% accuracy for most questions, with the lowest accuracy (for spiral arm count) being 98.6\%. The confusion matrices reflect this, showing little notable confusion for any question.

\begin{figure}
    \centering
    \includegraphics[width=\columnwidth]{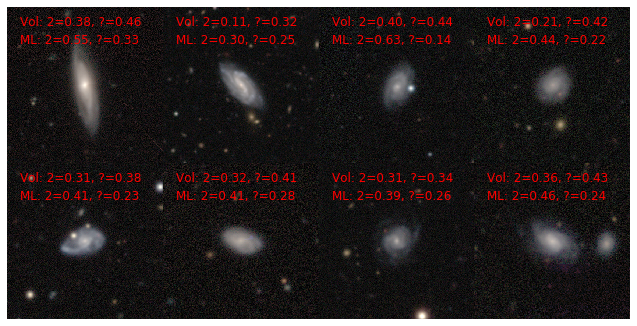}
    \includegraphics[width=\columnwidth]{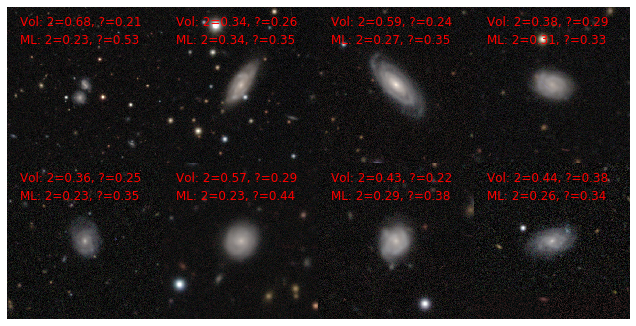}
    \caption{Random spiral galaxies where the classifier confuses the most likely volunteer vote for spiral arm count between `2' and `Can't Tell'. Above: galaxies where the classifier predicted `2' but more volunteers answered `Can't Tell'. Below: vice versa, galaxies where the classifier predicted `Can't Tell' but more volunteers answered `2'. Red text shows the volunteer (vol.) and machine-learning-predicted (ML) vote fractions for each answer. Counting the spiral arms is challenging, even for the authors. This highlights the difficulty in assessing performance by reducing the posteriors to classifications and then comparing against uncertain true labels.}
    \label{fig:confused_spirals}
\end{figure}

Reported performance on all galaxies will be lower than on confident galaxies as the correct labels are uncertain. Our measured vote fractions are approximations of the theoretical `true' vote fractions (as we cannot ask infinitely many volunteers), and many galaxies are genuinely ambiguous and do not have a meaningful `correct' answer. No classifier should achieve perfect accuracy on galaxies where the volunteers themselves are not confident. Nonetheless, performance is more than sufficient for scientific use; accuracy ranges from 77.4\% (spiral arm count) to 98.7\% (disk edge on). We observe some moderate confusion between similar answers, particularly between No or Weak bar, Moderate or Large bulges, and Two or Three spiral arms, which matches our intuition for the answers that volunteers might confuse and so likely reflects ambiguity in the training data. More surprisingly, there is also confusion between Two spiral arms and Can't Tell. Figure \ref{fig:confused_spirals} shows random examples of spirals where the most common volunteer answer was Two, but the classifier predicted Can't Tell, and vice versa. In both cases, the galaxies generally have diffuse or otherwise subtle spiral arms embedded in a bright disk, confusing both human and machine. This highlights the difficulty in using classification metrics to assess performance on ambiguous galaxies.

\begin{table}
    \begin{subtable}[h]{\columnwidth}
        \centering
        \begin{tabular}{l | l | l | l | l | l}
        \toprule
        Question & Count & Accuracy & Precision & Recall & F1 \\
        \midrule
        Smooth Or Featured & 11346 & 0.9352 & 0.9363 & 0.9352 & 0.9356 \\
        Disk Edge On & 3803 & 0.9871 & 0.9871 & 0.9871 & 0.9871 \\
        Has Spiral Arms & 2859 & 0.9349 & 0.9364 & 0.9349 & 0.9356 \\
        Bar & 2859 & 0.8185 & 0.8095 & 0.8185 & 0.8110 \\
        Bulge Size & 2859 & 0.8419 & 0.8405 & 0.8419 & 0.8409 \\
        How Rounded & 6805 & 0.9314 & 0.9313 & 0.9314 & 0.9313 \\
        Edge On Bulge & 506 & 0.9111 & 0.9134 & 0.9111 & 0.8996 \\
        Spiral Winding & 1997 & 0.7832 & 0.8041 & 0.7832 & 0.7874 \\
        Spiral Arm Count & 1997 & 0.7742 & 0.7555 & 0.7742 & 0.7560 \\
        Merging & 11346 & 0.8798 & 0.8672 & 0.8798 & 0.8511 \\
        
        \end{tabular}
       \caption{Classification metrics for all galaxies}
    \end{subtable}
    \newline
    \vspace*{0.25 cm}
    \newline
    \begin{subtable}[h]{\columnwidth}
        \centering
        \begin{tabular}{l | l | l | l | l | l}
        \toprule
        Question & Count & Accuracy & Precision & Recall & F1 \\
        \midrule
        Smooth Or Featured & 3495 & 0.9997 & 0.9997 & 0.9997 & 0.9997 \\
        Disk Edge On & 3480 & 0.9980 & 0.9980 & 0.9980 & 0.9980 \\
        Has Spiral Arms & 2024 & 0.9921 & 0.9933 & 0.9921 & 0.9924 \\
        Bar & 543 & 0.9945 & 0.9964 & 0.9945 & 0.9951 \\
        Bulge Size & 237 & 1.0000 & 1.0000 & 1.0000 & 1.0000 \\
        How Rounded & 3774 & 0.9968 & 0.9968 & 0.9968 & 0.9968 \\
        Edge On Bulge & 258 & 0.9961 & 0.9961 & 0.9961 & 0.9961 \\
        Spiral Winding & 213 & 0.9906 & 1.0000 & 0.9906 & 0.9953 \\
        Spiral Arm Count & 659 & 0.9863 & 0.9891 & 0.9863 & 0.9871 \\
        Merging & 3108 & 0.9987 & 0.9987 & 0.9987 & 0.9987 \\
        \end{tabular}
        \caption{Classification metrics for galaxies where volunteers are confident}
     \end{subtable}
     \caption{Classification metrics on all galaxies (above) or on galaxies where volunteers are confident for that question (i.e. where one answer has a vote fraction above 0.8). Multi-class precision, recall and F1 scores are weighted by the number of true galaxies for each answer. Classifications on confident galaxies are near-perfect.}. 
     \label{tab:ml_metrics}
\end{table}

\begin{figure}
    \centering
    \includegraphics[width=\columnwidth]{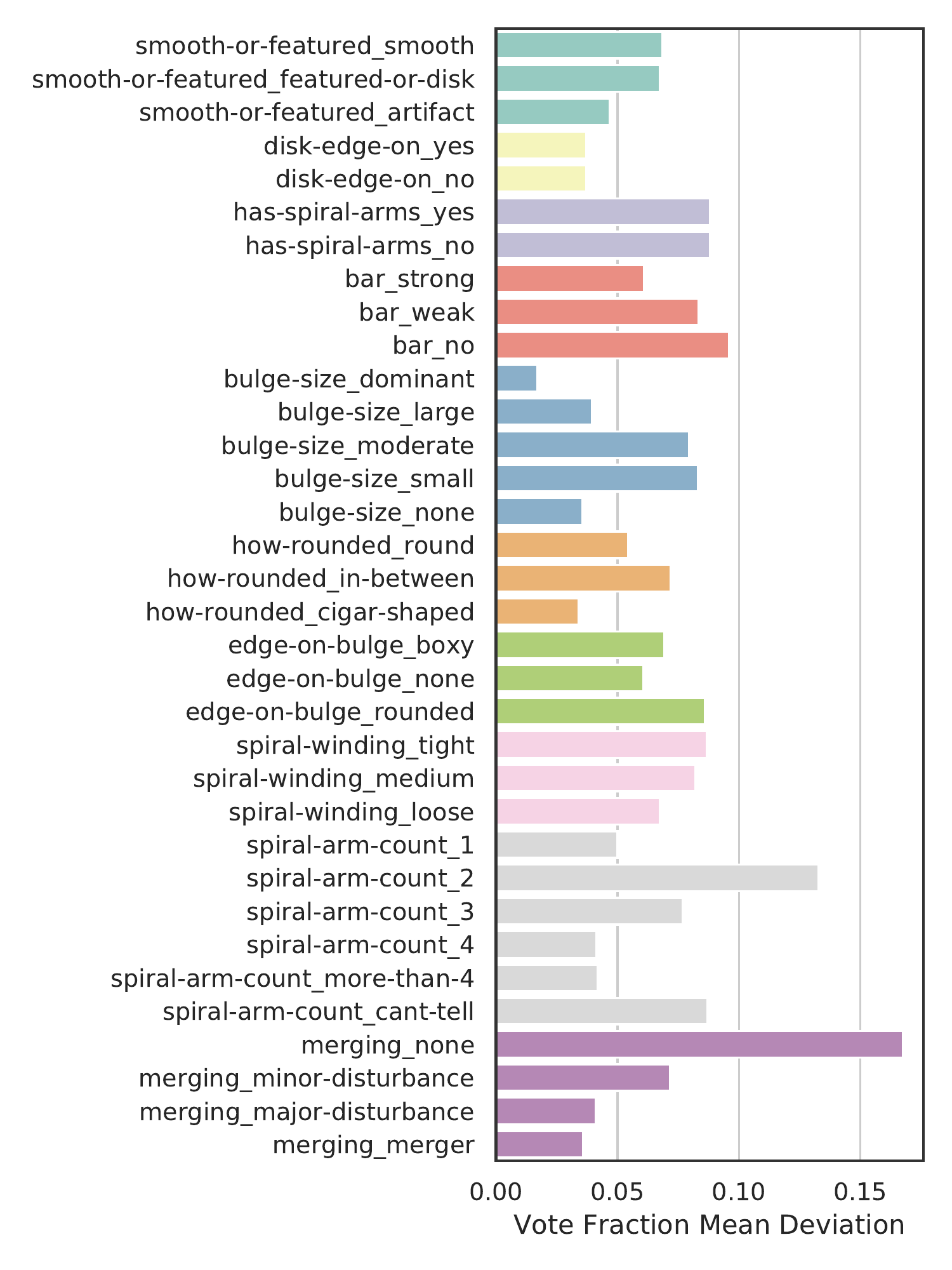}
    \caption{Mean absolute deviations between the model predictions and the observed vote fractions, by question, for the test set galaxies with approximately 40 volunteer responses. The model is typically well within 10\% of the observed vote fractions.}
    \label{fig:mean_deviation_bar}
\end{figure}

We can mitigate the ambiguity in classifications of galaxies by measuring regression metrics on the vote fractions, without rounding to discrete classifications. Figure \ref{fig:mean_deviation_bar} shows the mean deviations between the model predictions (mean posteriors) and the observed vote fractions, by question, for test set galaxies with approximately 40 volunteer responses. Performance is again excellent, with the predictions typically well within 10\% of the observed vote fractions. Predicting spiral arm count is relatively challenging, as noted above. Predicting answers to the `Merger' question of `None' (i.e. not a merger) is also challenging, perhaps because of the rarity of counter-examples.

\begin{figure}
    \centering
    \includegraphics[width=.9\columnwidth]{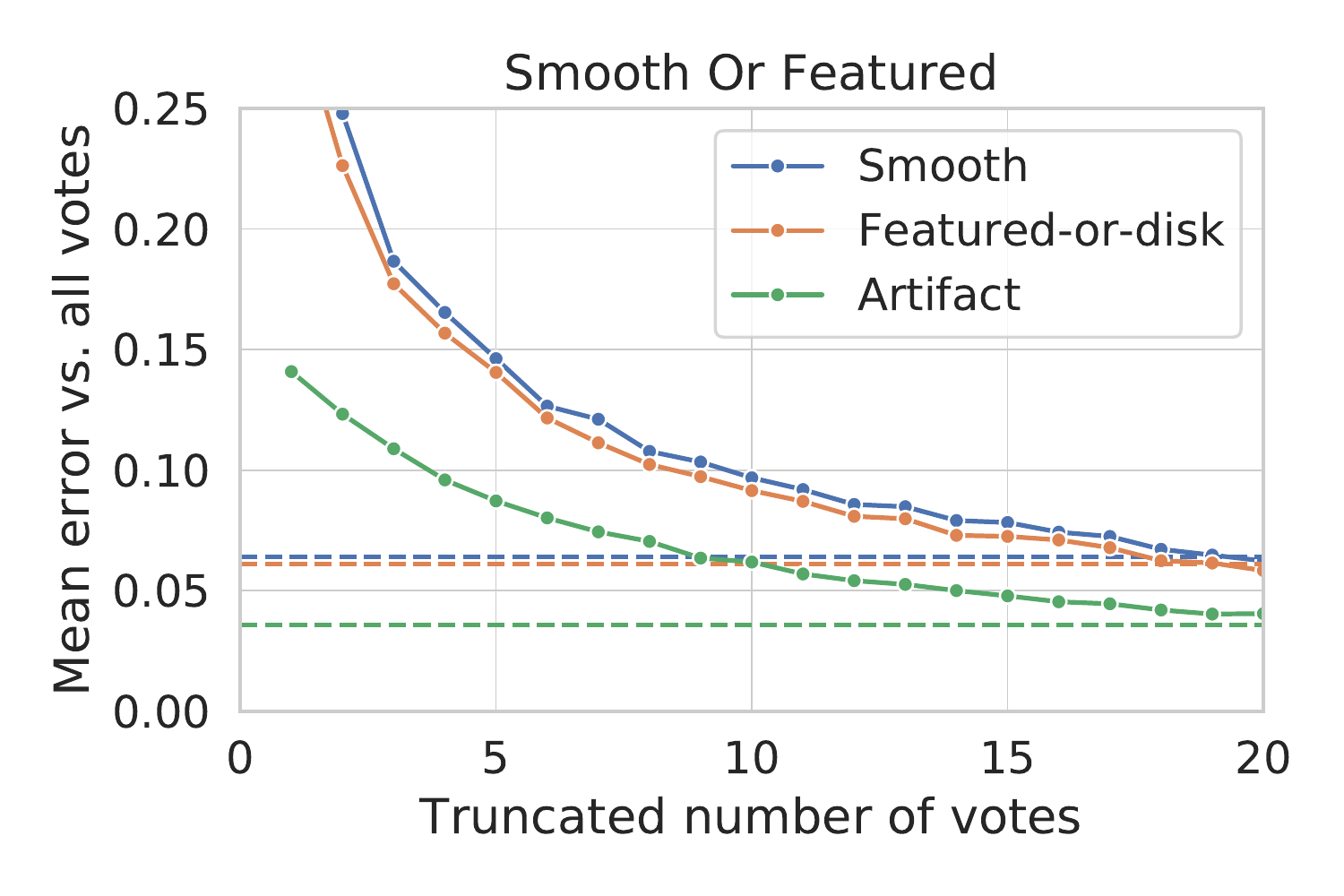}
    \includegraphics[width=.9\columnwidth]{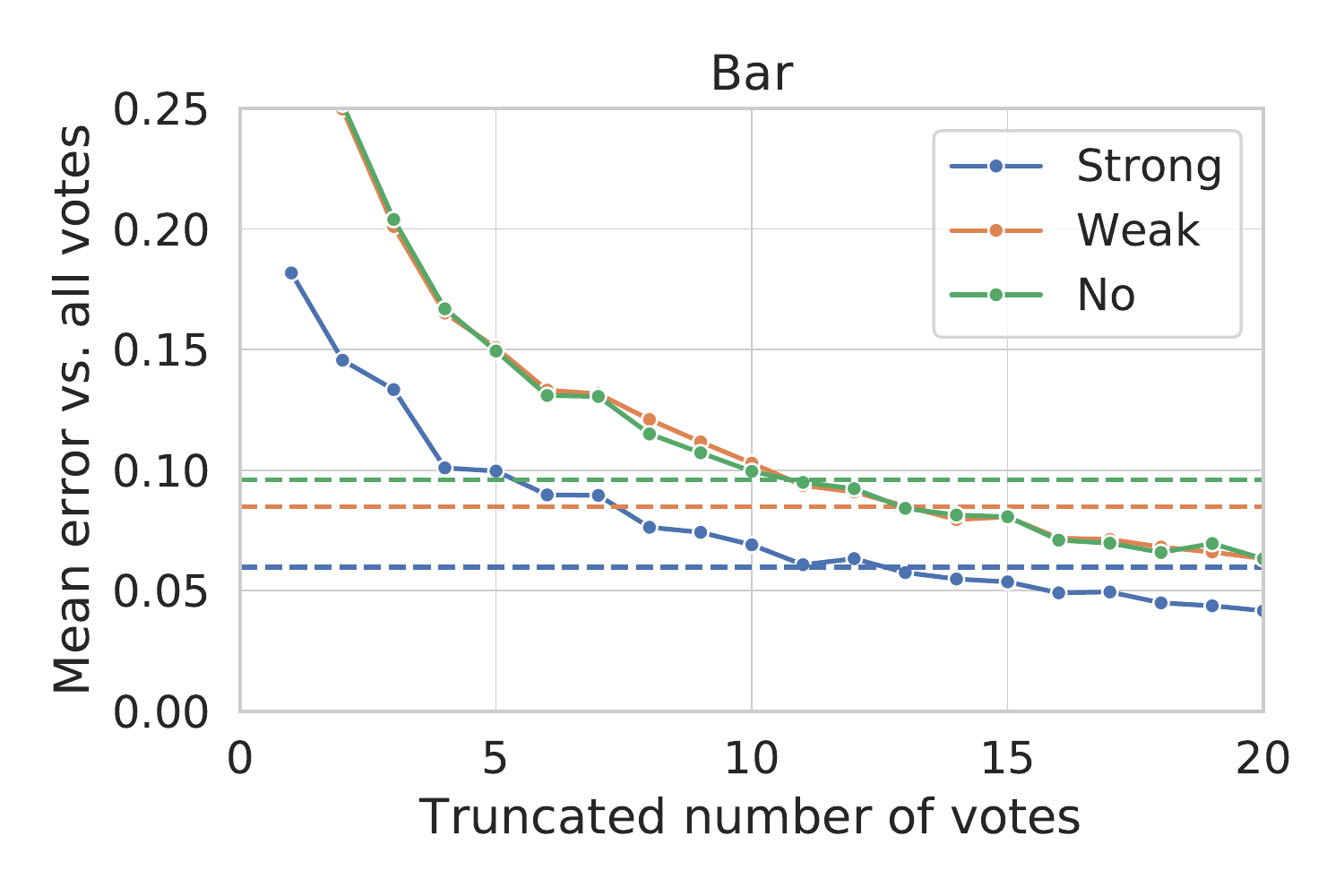}
    \includegraphics[width=.9\columnwidth]{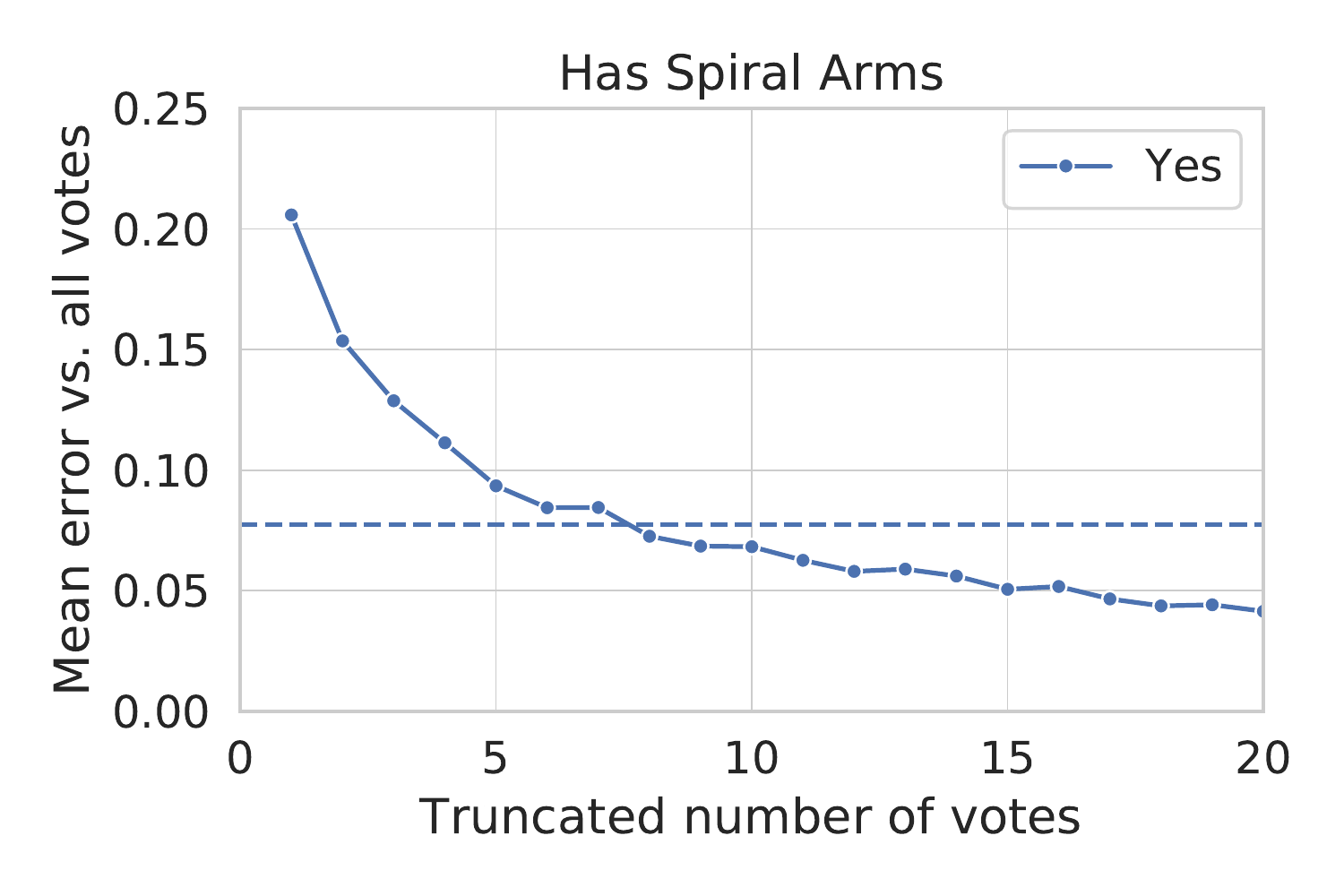}
    \includegraphics[width=.9\columnwidth]{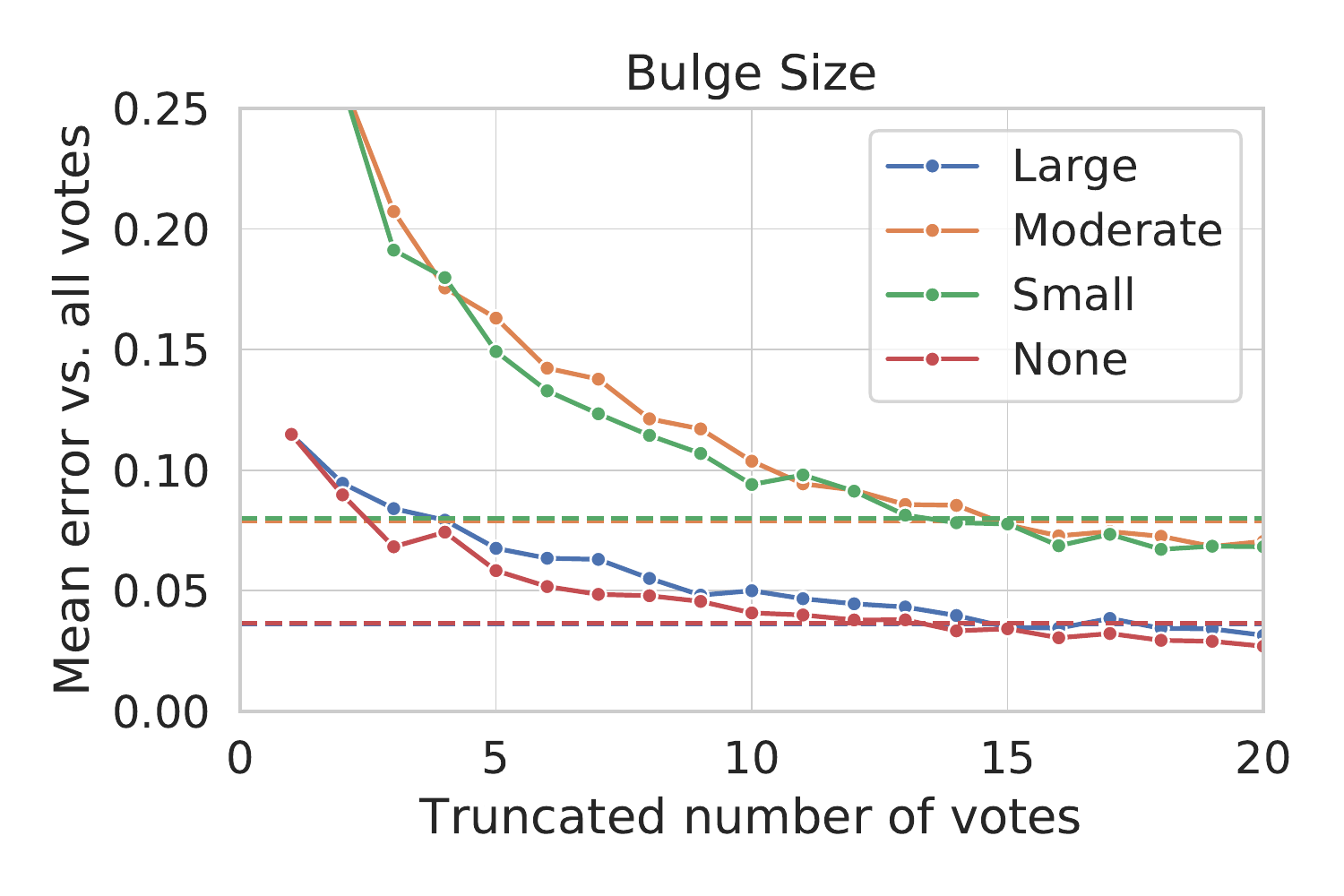}
    \caption{Mean error on the true ($N > 75$) vote fractions for either a truncated ($N=0$ to $N=20$) number of volunteers (solid) or the automated classifier (dashed). Asking only a few volunteers gives a noisy estimate of the true vote fraction. Asking more volunteers reduces this noise. For some number of volunteers, the noise in the vote fraction is similar to the error of the automated classifier, meaning they have a similar mean error vs. the true vote fraction; this number is where the solid and dashed lines intersect. We find the automated classifier has a similar mean error to approx. 5 to 15 volunteers, depending on the question.
    }
    \label{fig:deviation_vs_volunteers}
\end{figure}

The volunteer vote fractions against which we compare our predictions are themselves uncertain for most galaxies. We aim to predict the true vote fraction, i.e. the vote fraction from $\lim_{N \to \infty}$ volunteers, but we only know the vote fraction from $N$ volunteers. However, 387 pre-active-learning galaxies were erroneously uploaded twice or more, and so received more than 75 classifications each. The vote fractions for these $N > 75$ galaxies will be very similar to the $\lim_{N \to \infty}$ true vote fraction limit, allowing us to accurately measure the mean vote fraction error of our machine learning predictions. We can also calculate the mean vote fraction error (vs. the $N > 75$ vote fractions) from asking fewer ($N << 75$) volunteers by artificially truncating the number of votes collected, and ask - how many volunteer responses to that question would we need to have errors similar to that of our model? Note that the actual number of volunteers needed to be shown that galaxy to achieve an equivalent mean squared error will be higher for questions only asked given certain previous answers (i.e. all but `Smooth or Featured?' and `Merger?'), as some will give different answers to preceding questions and so not be asked that question. Figure \ref{fig:deviation_vs_volunteers} shows the model and volunteer mean errors for a representative selection of questions; the model predictions are as accurate as asking that question to around 10 volunteers.\footnote{The model is, in this strict sense, slightly superhuman.} 

\begin{figure}
    \centering
    \includegraphics[width=\columnwidth]{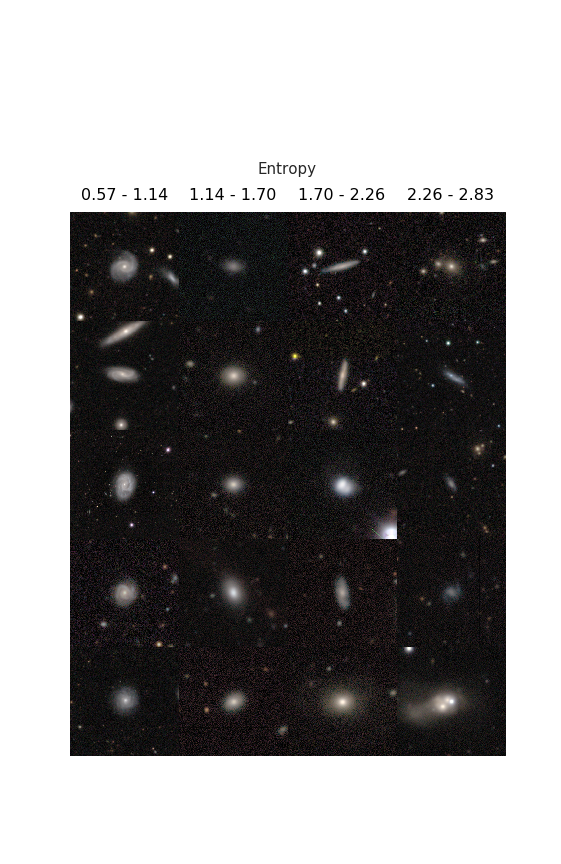}
    \caption{Galaxies binned by `Smooth or Featured' vote prediction entropy, measuring the model's uncertainty in the votes. Bins (columns) are equally spaced (boundaries noted above). Five random galaxies are shown per bin. Unusual, inclined or poorly-scaled galaxies have highly uncertain (high entropy) votes, while smooth and especially clearly featured galaxies have confident (low entropy) votes, matching our intuition and demonstrating that our posteriors provide meaningful uncertainties.}
    \label{fig:binned_uncertainty}
\end{figure}

We can also measure if our posteriors correctly estimate this uncertainty. As a qualitative test, Figure \ref{fig:binned_uncertainty} shows a random selection of galaxies binned by `Smooth or Featured' vote fraction prediction entropy, measuring the model's uncertainty. Prediction entropy is calculated as the (discrete) Shannon entropy $\sum_{\omega} p(\omega)\log(p(\omega))$ over all possible combinations of votes $\omega$, assuming 10 total votes for this question (our results are robust to other choices of total votes). Unusual, inclined or poorly-scaled galaxies have highly uncertain (high entropy) votes, while smooth and especially clearly featured galaxies have confident (low entropy) votes. The most uncertain galaxies (not shown) are so poorly scaled (due to incorrect estimation of the Petrosian radius in the NASA-Sloan Atlas) that they are barely visible. These results match our intuition and demonstrate that our posteriors provide meaningful uncertainties. 

\begin{figure}
    \centering
    \includegraphics[width=\columnwidth]{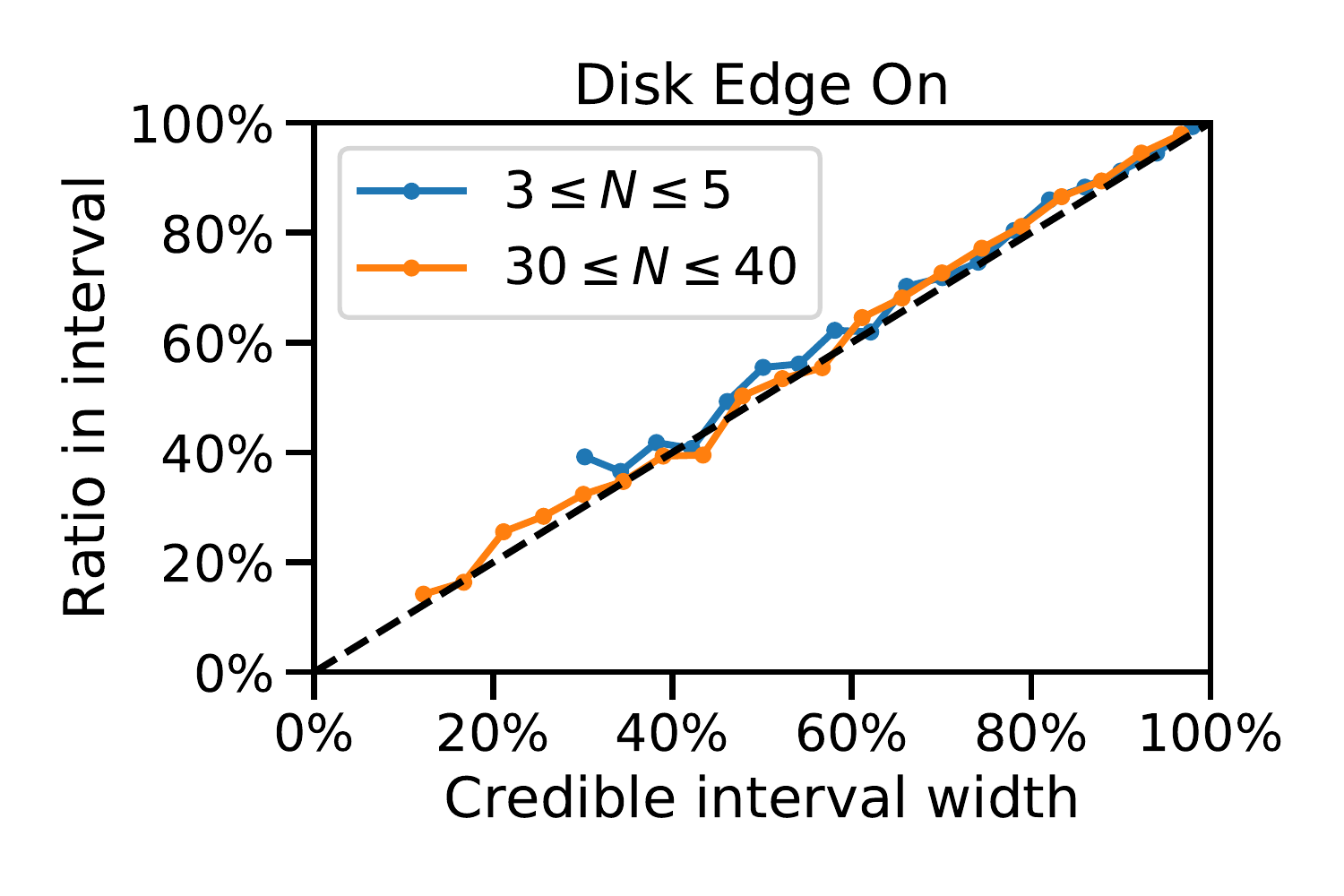}
    \includegraphics[width=\columnwidth]{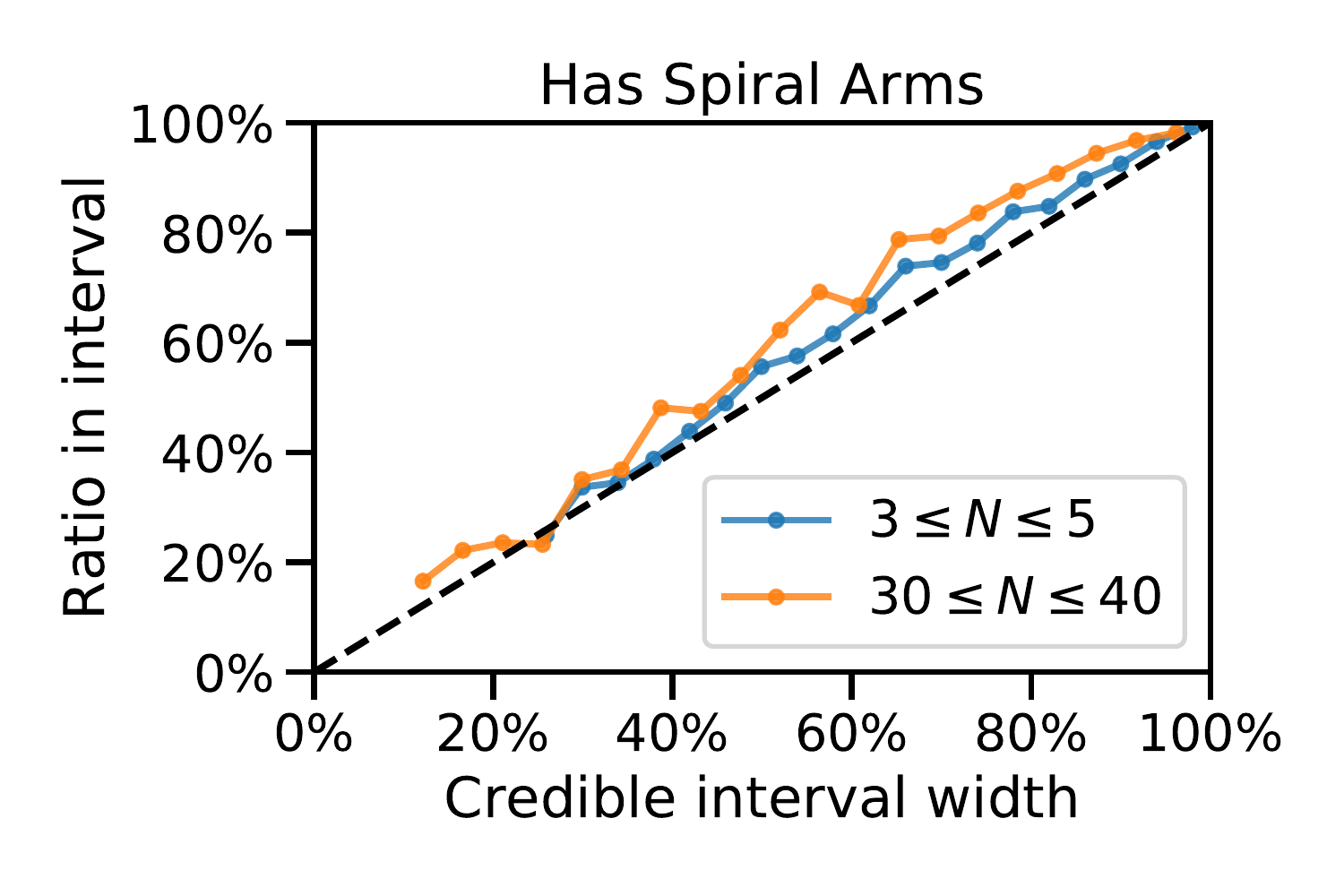}
    \caption{Calibration curves for the two binary GZ DECaLS questions. The $x$-axis shows the credible interval width - for data-dominated posteriors, roughly (e.g.) 30\% of galaxies should have vote fractions within their 30\% credible interval. The $y$-axis shows what percentage actually do fall within each interval width. We split calibration by galaxies with few votes (and hence typically wider posteriors) and more votes (narrower posteriors). Only credible intervals with at least 100 measurements are shown. Calibration for both questions is excellent.}
    \label{fig:calibration}
\end{figure}

More quantitatively, Figure \ref{fig:calibration} shows the calibration of our posteriors for the two binary questions in GZD-5 - `Edge-on Disk' and `Has Spiral Arms'. A well-calibrated posterior dominated by data (i.e. where the prior has minimal effect) will include the measured value within any bounds as often as the total probability within those bounds. We calculate calibration by, for each galaxy, iterating through each symmetric highest posterior density credible interval (i.e. starting from the posterior peak and moving the bounds outwards) and recording both the total probability inside the bounds and whether the recorded volunteer vote is inside the bounds. We then group (bin) by total probability and record the empirical frequency with which the votes lie within bounds of that total probability. In short, we are checking if, for all X, the observed value (vote fraction) falls within X\% of the posterior interval X\% of the time \citep{Cook2006, Levasseur2017}.
We find that calibration on these binary questions is excellent. Our classifier is correctly uncertain.

\begin{figure*}

    \includegraphics[width=.5\textwidth]{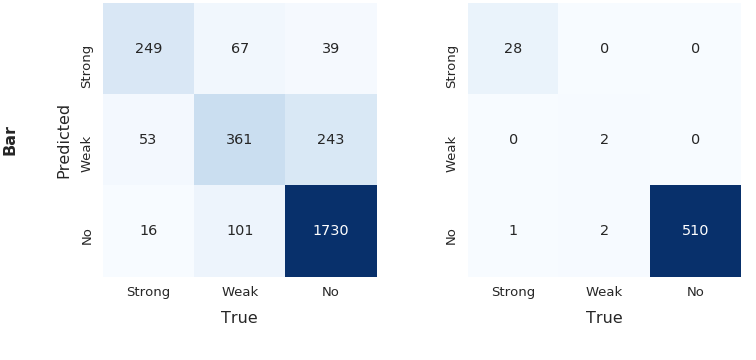}
        
    \includegraphics[width=.5\textwidth]{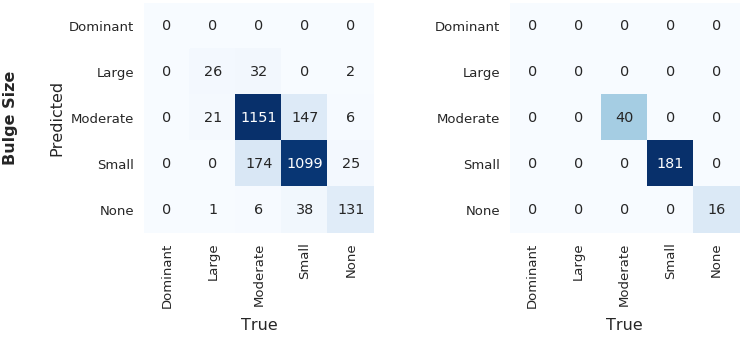}
    
    \includegraphics[width=.5\textwidth]{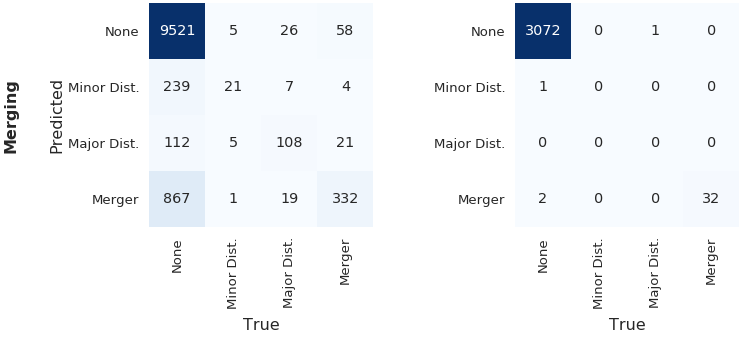}
    
    \includegraphics[width=.5\textwidth]{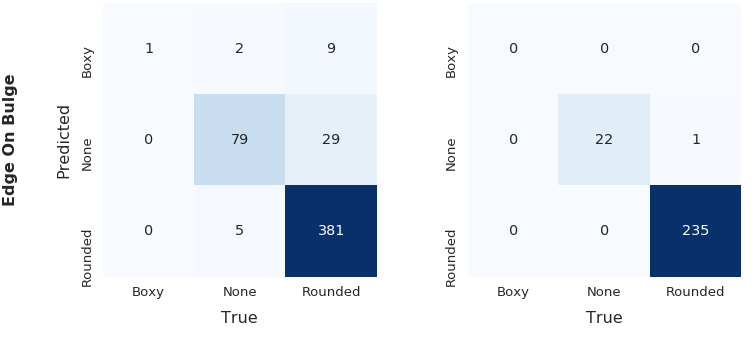}
    
    \includegraphics[width=.5\textwidth]{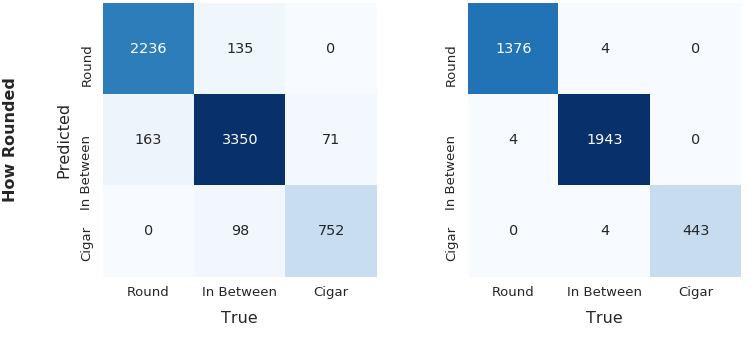}

    \caption{Confusion matrices for test set galaxies where the volunteers are confident in that question, defined as having the vote fraction for one answer above 0.8. Such confident galaxies are expected to have a clearly correct label, making correct and incorrect predictions straightforward to measure but also making the classification task easier. To avoid the loss of information from rounding, we encourage researchers not to treat Galaxy Zoo classifications as discrete, and instead to use the full vote fractions or posteriors where possible. }
\end{figure*}

The ultimate measure of success is whether our predictions are useful for science. \citealt{Masters2019} (hereafter M19) used GZ2 classifications to investigate the relationship between bulge size and winding angle and found - contrary to a conventional view of the Hubble sequence - no strong correlation. We repeat this analysis using our (deeper) DECaLS data, using either volunteer or automated classification, to check if the automated classifications lead to the same science results as the volunteers.

Specifically, we select a clean sample of face-on spiral galaxies using M19's vote fraction cuts of $f_{\text{feat}} > 0.43$, $f_{\text{not-edge-on}} > 0.715$, and $f_{\text{spiral-yes}} > 0.619$. We also make a cut of $f_{\text{merging=none}} > 0.5$, analogous to M19's $f_{\text{odd}}$ cut, to remove galaxies with ongoing mergers or with otherwise disturbed features. For the volunteer vote fractions, we can only use either GZD-1/2 or GZD-5 classifications, since the former decision tree had three bulge size answers and the latter had five; we choose GZD-5 to benefit from the added precision of additional answers. To avoid selection effects (Sec. \ref{sec:selection_effects}) we only use galaxies classified prior to active learning being activated. For the automated classifications, we use a model trained on GZD-5 to predict GZD-5 decision tree vote fractions (including the five bulge answers) for every GZ DECaLS galaxy (313,798). This allows us to expand our sample size from 5,378 galaxies using GZD-5 volunteers only to 43,672 galaxies using our automated classifier.

We calculate bulge size and spiral winding following Eqn. 1 and 3 in M19, trivially generalising the bulge size calculation to allow for five bulge size answers:

\begin{gather}
    \label{eqn:summary_stats}
    W_{\text{avg}} = 0.5 f_{\text{medium}} + 1.0 f_{\text{tight}}  \\
    B_{\text{avg}} = 0.25 f_{\text{small}} + 0.5 f_{\text{moderate}} + 0.75 f_{\text{large}} + 1.0 f_{\text{dominant}}
\end{gather}

\begin{figure}
    \centering
    \includegraphics[width=0.49\columnwidth]{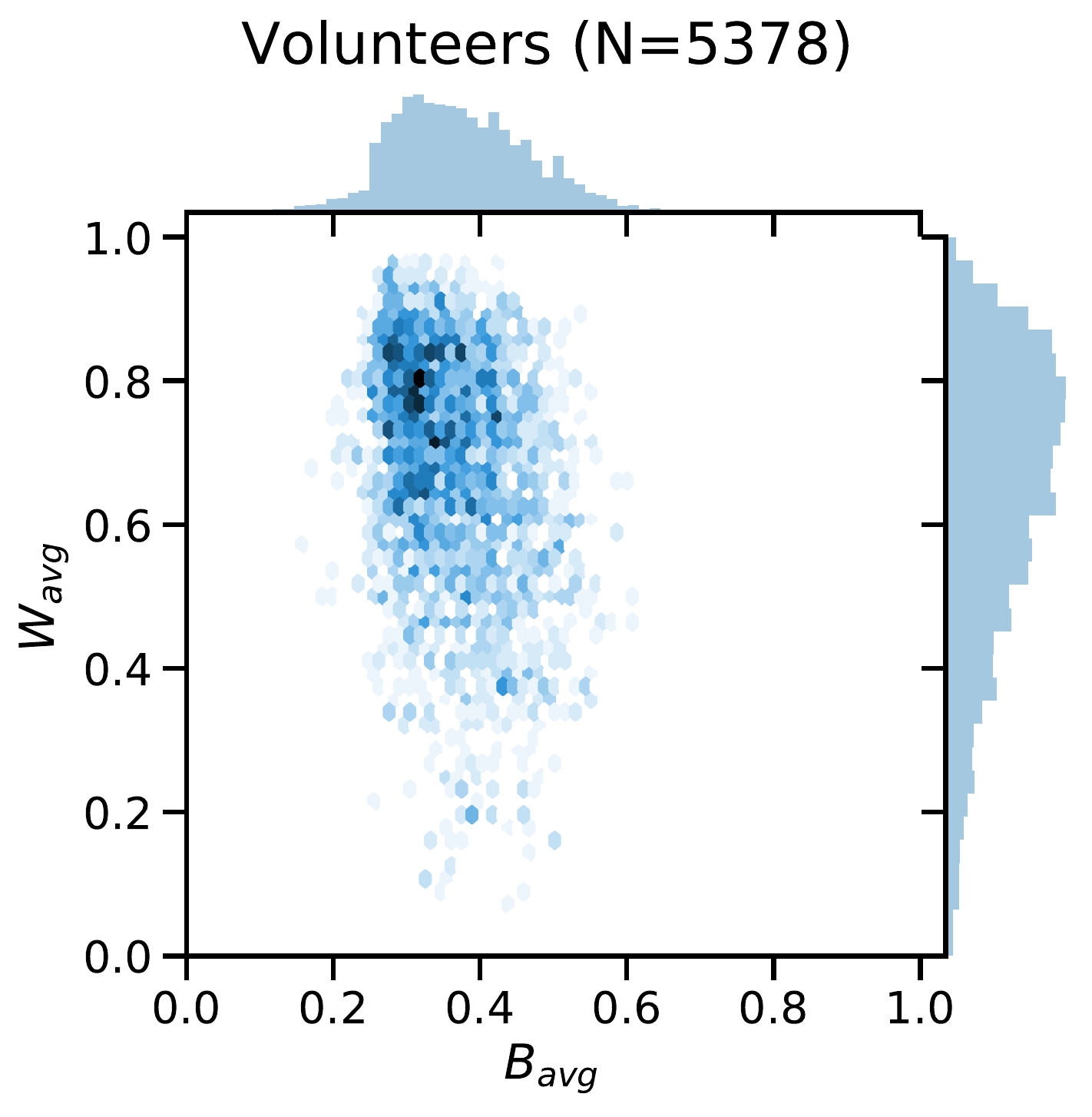}
    \includegraphics[width=0.49\columnwidth]{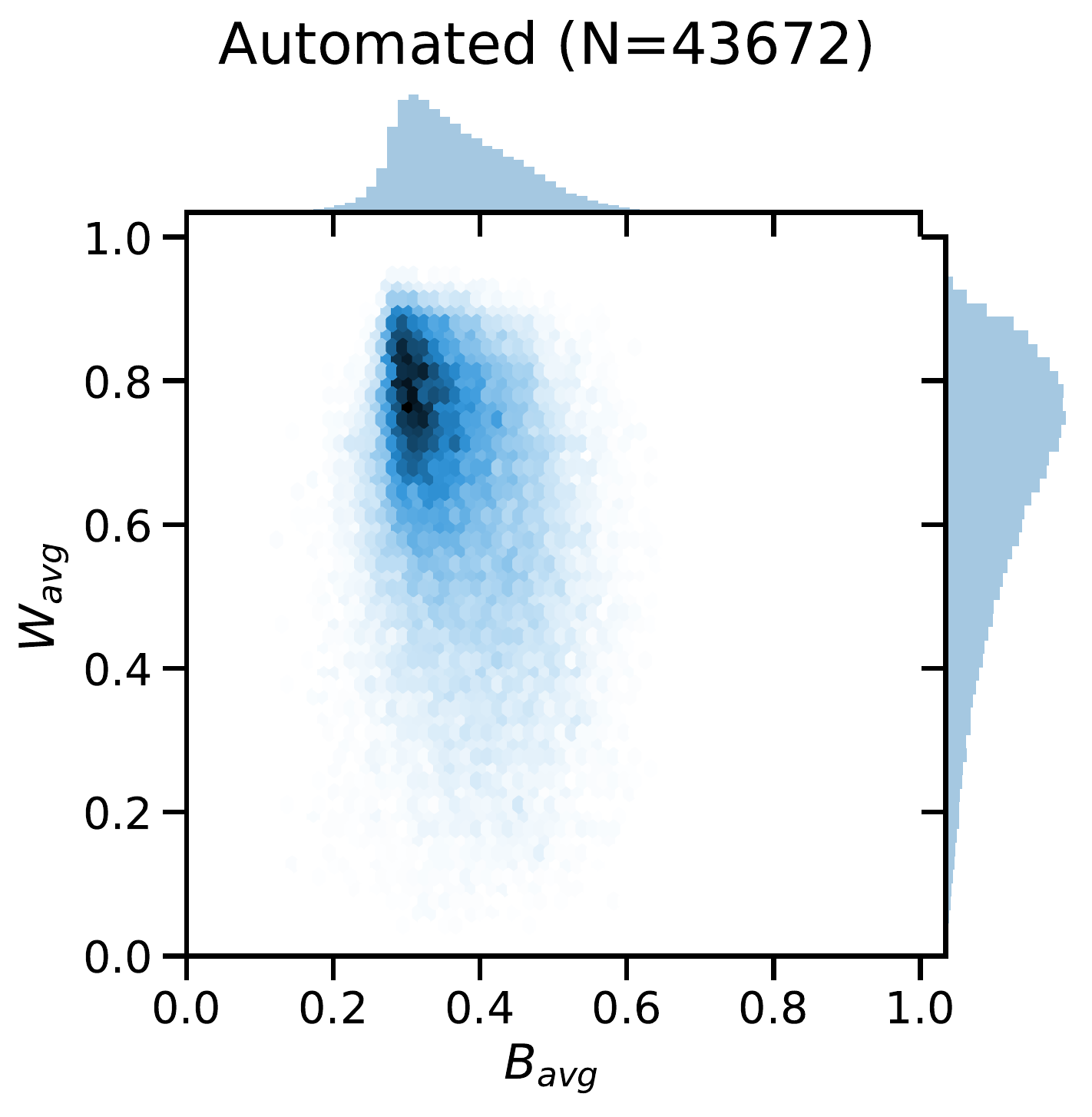}
    \caption{Distribution of bulge size vs. spiral winding, using responses from volunteers (left) or our automated predictions (right). We observe no clear correlation between bulge size and spiral winding, consistent with M19. The distributions are consistent between volunteers and our automated method. We hope this demonstrates the accuracy and scientific value of our automated classifier.}
    \label{fig:masters_repeat}
\end{figure}

Both classification methods find no correlation between bulge size and spiral winding, consistent with M19. Figure \ref{fig:masters_repeat} shows the distribution of bulge size against spiral winding using either volunteer predictions (fractions) or the deep learning predictions (expected fractions) for the sample of featured face-on galaxies selected above. The distributions are indistinguishable, with the automated method offering a substantially larger (approx 8x) sample size. We hope this demonstrates the accuracy and scientific value of our automated classifier.

\section{Usage}
\label{sec:usage}

\subsection{Catalogues}

We release two volunteer catalogues and two automated catalogues, available at \href{https://doi.org/10.5281/zenodo.4573248}{https://doi.org/10.5281/zenodo.4573248}.

\texttt{gz\_decals\_volunteers\_ab} includes the volunteer classifications for 92,960 galaxies from GZD-1 and GZD-2. Classifications are made using the GZD-1/2 decision tree (Fig. \ref{fig:gzda_tree}). All galaxies received at least 40 classifications, and consequently have approximately 30-40 after volunteer weighting (Sec. \ref{sec:volunteer_weighting}). This catalogue is ideal for researchers needing standard morphology measurements on a reasonably large sample, with minimal complexity. 33,124 galaxies in this catalogue were also previously classified in GZ2; the GZD-1/2 classifications are better able to detect faint features due to deeper DECaLS imaging, and so should be preferred.

\texttt{gz\_decals\_volunteers\_c} includes the volunteer classifications from GZD-5. Classifications are made using the improved GZD-5 decision tree which adds more detail for bars and mergers (Sec. \ref{sec:comparison_of_decision_trees}). This catalogue includes 253,286 galaxies, but each galaxy does not have the same number of classifications. 59,337 galaxies have at least 30 classifications (after denoising), and the remainder have far fewer (approximately 5). The selection effects for how many classifications each galaxy receives are detailed below in Sec. \ref{sec:selection_effects}. This catalogue may be useful to researchers who prefer a larger sample than  \texttt{gz\_decals\_volunteers\_ab} at the cost of more uncertainty and the introduction of selection effects, or who need detailed bar or merger measurements for a small number of galaxies. We use \texttt{gz\_decals\_volunteers\_c} to train our deep learning classifier.

The automated classifications are made using our Bayesian deep learning classifier, trained on \texttt{gz\_decals\_volunteers\_c} to predict the answers to the GZD-5 decision tree for all GZ DECaLS galaxies (including those in GZD-1 and GZD-2). \texttt{gz\_decals\_auto\_posteriors} contains the predicted posteriors for each answer - specifically, the Dirichlet concentration parameters that encode the posteriors. We hope this catalogue will be helpful to researchers analysing galaxies in Bayesian frameworks.

\texttt{gz\_decals\_auto\_fractions} reduces those posteriors to the automated equivalent of previous Galaxy Zoo data releases, containing the expected vote fractions (mean posteriors). Note that not all vote fractions are relevant for every galaxy; we suggest assessing relevance using the estimated fraction of volunteers that would have been asked each question, which we also include. We hope this catalogue will be useful to researchers seeking detailed morphology classifications on the largest possible sample, who might benefit from error bars but do not need full posteriors.

We also release Jupyter notebooks showing how to use each catalogue on \url{www.github.com} (full link on publication). These demonstrate how to load and query each catalogue with \texttt{pandas} \citep{McKinney2010}, and how to create callable posteriors from the Dirichlet concentration parameters.

The automated catalogues may be interactively explored at \href{https://share.streamlit.io/mwalmsley/galaxy-poster/gz_decals_mike_walmsley.py}{https://share.streamlit.io/mwalmsley/galaxy-poster/gz\_decals\_mike\_walmsley.py}.

\subsection{Selection Effects for Total Classifications}
\label{sec:selection_effects}
The GZD-1/2 catalogue reports at least 40 classifications for all galaxies imaged by DECaLS DR1/2 and passing the appropriate selection cuts (Section \ref{sec:selection}). Additional classifications above 40 are assigned independently of the galaxy properties. The selection function for total classifications in the GZD-5 catalogue is more complex. In practice, if you require a strictly random sample of GZD-5 galaxies with more than five volunteer classifications, you should exclude galaxies where `random\_selection' is False. You may also consider using the posteriors from our deep learning classifier, which are comparable across all GZ DECaLS galaxies (Section \ref{sec:automated}). Below, we describe the GZD-5 total classification selection effects. 

Early galaxies were initially uploaded row-by-row from the NASA-Sloan Atlas, each (eventually) receiving 40 classifications. We also uploaded two additional subsets. For the first, 1355 galaxies were targeted for classification to support an external research project. Of these, 1145 would have otherwise received five classifications. These 1145 galaxies with additional classifications are identified with the `targeted' group and should be excluded. For the second, we reclassified the 1497 galaxies classified in both GZD-1/2 and the \cite{Nair2010} expert visual morphology classification catalogue to measure the effect of our new decision tree (results are shown in Sec. \ref{sec:comparison_of_decision_trees}). Both the GZD-1/2 and GZD-5 classifications are reported in the respective catalogues (Section \ref{sec:usage}. Similarly to the targeted galaxies, 651 of these calibration galaxies would have otherwise received five classifications, are identified with the `calibration' group, and should be excluded.

We then implemented active learning (Sec \ref{sec:retirement}), prioritising 6,939 galaxies from the remaining pool of 199,496 galaxies not yet uploaded. The galaxies are identified with the groups `active\_priority' (the galaxies identified as `most informative' and selected for 40 classifications) and `active\_baseline' (the remainder). For a strictly random selection, both groups should be excluded, leaving the galaxies classified prior to the introduction of active learning.

Finally, we note that 14,960 (5.9\%) of GZD-5 galaxies received more than 40 classifications due to being erroneously uploaded more than once. The images are identical and so we report the aggregate classifications across all uploads of the same galaxy.

\subsection{Suggested Usage of Vote Fractions}

The most appropriate usage of the Galaxy Zoo DECaLS vote fractions depends on the specific science case. Many galaxies have ambiguous vote fractions (e.g. roughly similar vote fractions for both disk and elliptical morphologies) because of observational limitations like image resolution, or because the galaxy morphology is truly in-between the available answers (perhaps because the galaxy has an unusual feature such as polar rings, \citealt{Moiseev2011}, or because the galaxy is undergoing a morphological transition). To make best use of such galaxies, we recommend that, where possible, readers use the vote fractions as statistical weights in their analysis. For example, when investigating the differences in the stellar mass distributions of elliptical and disk galaxies, the disk (elliptical) vote fractions can be used as weights when plotting the distributions, resulting in the galaxies with the highest vote fraction for disk (elliptical) morphology dominating the resulting distribution. This ensures that each galaxy contributes to the analysis, without excluding galaxies with ambiguous vote fractions. For examples of using vote fractions as weights, see \cite{Smethurst2015} and \cite{Masters2019}.

Using the vote fractions as weights is not appropriate for all science cases. For example, if galaxies of a particular morphology need to be isolated to form a sample for observational follow-up (e.g. overlapping pairs, see \citealt{Keel2013}, and `bulgeless' galaxies, see \citealt{Simmons2017a,Smethurst2019}), or if the fraction of a certain morphological type of galaxy is to be calculated \citep[e.g. bar fraction, see][]{Simmons2014}. These science cases require a cut on the appropriate vote fraction to be chosen. However, readers should be aware that making cuts on the vote fractions is a crude method to identify galaxies of certain morphologies and will result in an incomplete sample.

Table \ref{tab:suggested_cuts} shows our suggested cuts for populations of common interest, based on visual inspection by the authors and chosen for high specificity (low contamination) at the cost of low sensitivity (completeness). We urge the reader to adjust these cuts to suit the sensitivity and specificity of their science case, to add additional cuts to better select their desired population, and to make their own visual inspection to verify the selected population is as intended. For a full analysis, we once again suggest the reader avoid cuts by appropriately weighting ambiguous galaxies, or take advantage of the posteriors provided by our automated classifier. 

\begin{table*}
    \centering
    \begin{tabular}{ |p{2.5cm}||p{4cm}|p{1.4cm} |p{4cm}}
    
        \hline
        Population & Approx. Cut & Q. Votes & Notes\\
        \hline
        Featured Disk   & $\texttt{featured} > 0.7$    & 5 & \\
        Disk   & $\texttt{featured} > 0.3$    & 5 & Will include featureless S0  \\
        Elliptical   & $\texttt{smooth} > 0.7$    & 5 & \\
        \hline
        Edge-on Disk   & $\texttt{yes} > 0.8$    & 5 & \\
        Not Edge-on Disk   & $\texttt{yes} < 0.3$    & 5 & \\
        \hline
        Strong Bar & $\texttt{strong bar} > 0.8$ & 20 & \\
        Weak Bar & $\texttt{weak bar} > 0.8$ & 20 & \\
        Any Bar & $\texttt{strong bar} + \texttt{weak bar} > 0.6$ & 20 & \\
        \hline
        Spiral Arms & $\texttt{spiral arms} > 0.6$ & 20 & \\
        No Spiral Arms & $\texttt{spiral arms} < 0.3$ & 20 & Primarily ringed or irregular\\
        \hline
        Spiral Count & $\texttt{spiral count \{n\}} > 0.75$ & 30 & One-armed spirals are often mergers \\
        \hline
        Round Edge-on Bulge & $\texttt{edge-on bulge rounded} > 0.6$ & 10 & \\
        Boxy Edge-on Bulge & $\texttt{edge-on bulge boxy} > 0.3$ & 10 & Rare - visual inspection required\\
        No Edge-on Bulge & $\texttt{edge-on bulge none} > 0.5$ & 10 & \\
        \hline
        Merger   & $\texttt{merger} > 0.7$    & 10 & \\
        Merger or Overlap & $\texttt{merger} > 0.3$  & 10 & To remove overlaps, redshifts or inspection required. \\
        Post-Merger & $\texttt{major disturb.} > 0.6$ & 10 & \\
        Asymmetric or Low Surface Brightness & $\texttt{minor disturb.} > 0.4$ & 10 & \\
    \end{tabular}
    \caption{Suggested cuts for \textit{rough} identification of galaxy populations, based on visual inspection by the authors. Q. votes is the minimum number of total votes for that question; for example, to identify strong bars, require at least 20 total votes to the question `Does this galaxy have a bar?'. This ensures enough votes to calculate reliable vote fractions. Assumes that all previous questions are filtered with the suggested cuts. For continuous measurements such as bulge size and spiral winding, we suggest combining all answers into a summary statistic like Eqn. \ref{eqn:summary_stats}.}
    \label{tab:suggested_cuts}
\end{table*}

\section{Discussion}

What does a classification mean? The comparison of GZ2 and GZ DECaLS images (Fig. \ref{fig:featured_galaxies_big_shift}) highlights that our classifications aim to characterise the clear features of an image, and not what an expert might infer from that image. For example, volunteers might see an image of a galaxy that is broadly smooth, and so answer smooth, even though our astronomical understanding might suggest that the faint features around the galaxy core are likely indicative of spiral arms that would be revealed given deeper images. This situation occurs in several galaxies in Fig. \ref{fig:featured_galaxies_big_shift}. These `raw' classifications will be most appropriate for researchers working on computer vision or on particularly low-redshift, well-resolved galaxies. The redshift-debiased classifications, which are effectively an estimate of galaxy features \textit{not clearly seen} in the image, will be most appropriate for researchers especially interested in fainter features or studying links between our estimated intrinsic visual morphologies and other galaxy properties.

We showed in Sec. \ref{sec:comparison_of_decision_trees} that changing the answers available to volunteers significantly improves our ability to identify weak bars. This highlights that our classifications are only defined in the context of the answers presented. One cannot straightforwardly compare classifications made using different decision trees. Our scientific interests and our understanding of volunteers both evolve, and so our decision trees must also evolve to match them. However, only the last few years of volunteer classifications will use the latest decision tree (based on previous data releases), placing an upper limit on the number of galaxies with compatible classifications at any one time. Our automated classifier resolves this here by allowing us to retrospectively apply the GZD-5 decision tree (with better weak bar detection, among other changes) to galaxies only classified by volunteers in GZD-1 and GZD-2. This flexibility ensures that Galaxy Zoo will remain able to answer the most pertinent research questions at scale.

We have shown (\ref{sec:automated_results}) that our automated classifier is generally highly accurate, well-calibrated, and leads to at least one equivalent science result. However, we cannot exclude the possibility of unexpected systematic biases or of adversarial behaviour from particular images. Avoiding subtle biases and detecting overconfidence on out-of-distribution data remain open computer science research questions, often driven by important terrestrial applications \citep{Szegedy2014,Hendrycks2016,Eykholt2017,Smith2018,Geirhos2019,Ren2019,Yang2020,Margalef-Bentabol2020}. Volunteers also have biases (e.g a slight preference for recognising left-handed spirals, \citealt{Land2008}) and struggle with images of an adversarial nature (e.g. confusing edge-on disks with cigar-shaped ellipticals), though these can often be discovered and resolved through discussion with the community and by adapting the website.

We believe the future of morphology classification is in the thoughtful combination of volunteers and machine learning. Such combinations will be more than just faster; they will be replicable, uniform, error-bounded, and quick to adapt to new tasks. They will let us ask new questions - draw the spiral arms, select the bar length, separate the merging galaxies pixelwise - which would be infeasible with volunteers alone for all but the smallest samples (e.g. \citealt{Lingard2020}). And they will find the interesting, unusual and unexpected galaxies which challenge our understanding and inspire new research directions.

The best combination of volunteer and machine is unclear. Our experiment with active learning is one possible approach, but (when compared to random selection) suffers from complexity to implement, an unknown selection function, and no guarantee - or even clear final measurement of - an improvement in model performance. Many other approaches are suggested in astrophysics \citep{Wright2017,Beck2018,Wright2019,Dickinson2019,Martin2020,Lochner2020} and in citizen science and human-computer interaction more broadly \citep{Chang2017,Wilder2020,Liu2020,Bansal2019}. We will continue to search for and experiment with strategies to create the most effective contribution to research by volunteers.

\section{Conclusion}

We have presented Galaxy Zoo DECaLS; detailed galaxy morphology classifications for 311,488 galaxies imaged by DECaLS DR5 and within the SDSS DR11 footprint. The increased depth of DECaLS imaging allows us to better resolve faint morphological features than with previous Galaxy Zoo data releases using SDSS imaging (Fig. \ref{fig:featured_comparison}). Classifications were collected from volunteers on the Zooniverse citizen science platform over three campaigns, GZD-1, GZD-2, and GZD-5. GZD-5 used an improved decision tree (Fig. \ref{fig:decision_tree}) aimed at best exploiting the deeper DECaLS images to identify weak bars, mergers, and tidal features. 

All galaxies receive at least five volunteer classifications (Fig. \ref{fig:classification_counts}). Galaxies in GZD-1 and GZD-2 receive at least 40. In GZD-5, two subsets receive 40: a random subset and a subset of galaxies prioritised as most likely to be informative for training machine learning models. These informative galaxies were identified following the method introduced by \cite{Walmsley2020} as the galaxies with the highest mutual information between model parameters and volunteer labels - intuitively, the galaxies on which several machine learning models confidently disagree.

Volunteer classifications were then used to train deep learning models to classify all galaxies. Our models are able to both learn from uncertain volunteer responses and predict full posteriors (rather than point estimates) for what volunteers would have said. This was achieved by interpreting the model predictions as the parameters for Dirichlet-Multinomial distributions and training to maximise the corresponding likelihood (Eqn. \ref{multivariate_per_q_likelihood}). We also approximate marginalising over model weights (i.e. Bayesian deep learning) by training an ensemble of 5 models where each model makes predictions with MC Dropout. 
The resulting ensemble is accurate (Fig. \ref{fig:confusion_matrices}-\ref{fig:deviation_vs_volunteers}) and well-calibrated (Fig. \ref{fig:calibration}). 

We release both volunteer and automated classification catalogues at \href{data.galaxyzoo.org}{data.galaxyzoo.org}. The volunteer catalogues include the total and mean volunteer responses for each question to each galaxy, and are split into the GZD-1/2 and GZD-5 campaigns (due to the modified decision tree). The automated catalogue includes predictions for every galaxy in any campaign. We share the predicted Dirichlet-Multinomial parameters that encode the full posteriors as well as the expected vote fractions that those posteriors imply. The expected vote fractions may used in a similar manner to previous volunteer-only data releases, while the posteriors support more complex statistical analyses. We also provide guidance and code examples.

\appendix

\section{GZD-1/2 Decision Tree}
\label{sec:decision_trees}

\begin{figure*}
    \centering
    \includegraphics[width=\textwidth]{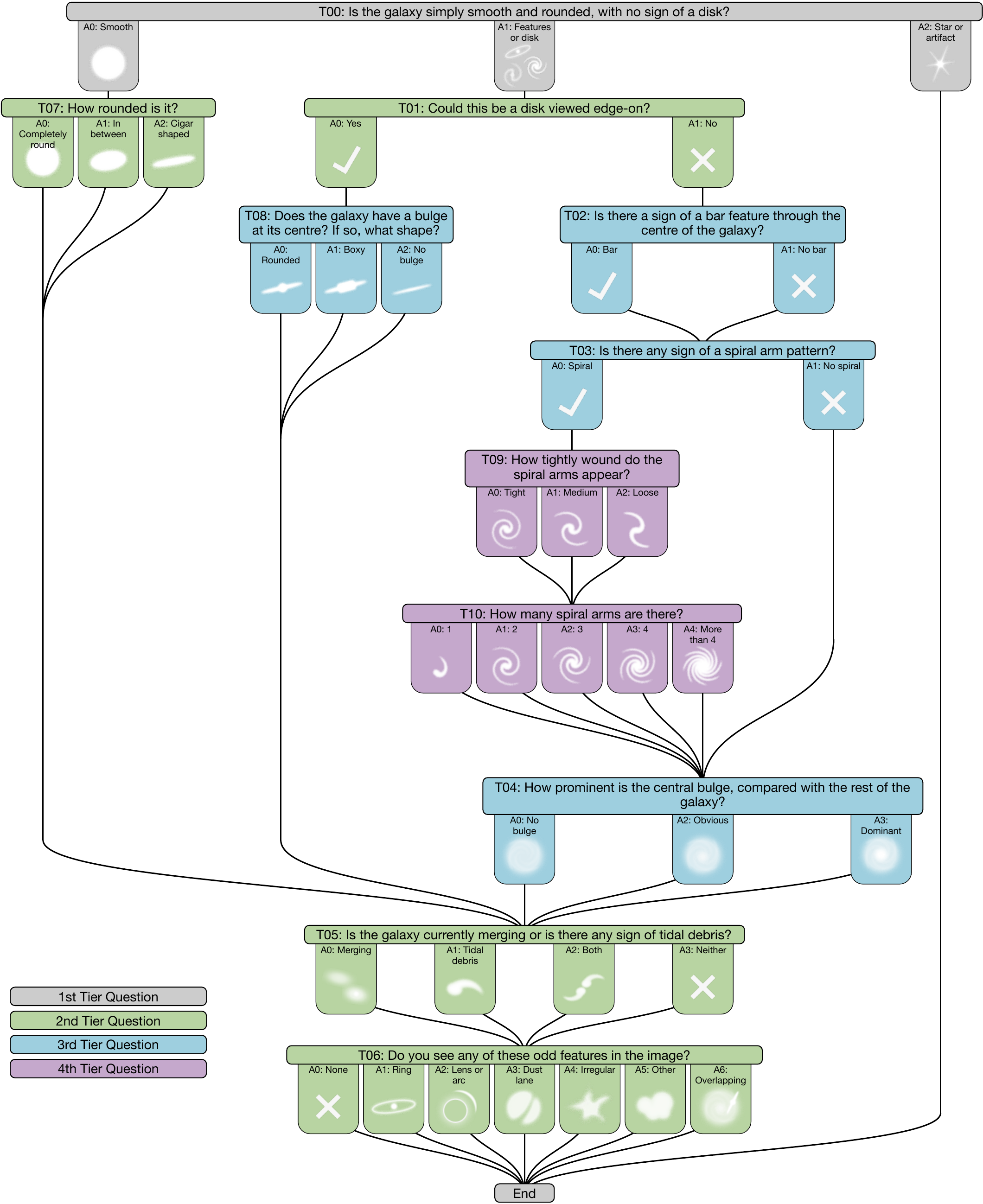}
    \caption{Decision tree used for GZD-1 and GZD-2, based on the Galaxy Zoo 2 decision tree. The GZD-5 decision tree is shown in Figure \ref{fig:decision_tree}.}
    \label{fig:gzda_tree}
\end{figure*}

Figure \ref{fig:gzda_tree} shows the Galaxy Zoo decision tree used for the earlier GZD-1 and GZD-2 DECaLS campaigns. This tree is based on the tree used for Galaxy Zoo 2 \citep{Willett2013} with three modifications; the `Can't Tell' answer to `How many spiral arms are there?' was removed, the number of answers to `How prominent is the central bulge?' was reduced from four to three, and `Is the galaxy currently merging, or is there any sign of tidal debris?' was added as a standalone question. Please see Sec. \ref{sec:decision_trees_intro} for a full discussion.

\section{Catalogue Sample Rows}

Tables \ref{tab:sample_volunteers} and \ref{tab:sample_ml} present sample rows from the volunteer and automated morphology catalogues respectively. The volunteer data shown is from GZD-5; the GZD-1/2 catalogue follows an equivalent schema. For brevity, we show only columns for a single question (`Bar') and a single answer (`Weak'); other questions and answers follow an identical pattern. A full description of all columns is available on \href{data.galaxyzoo.org}{data.galaxyzoo.org}.

\begin{table*}
    \centering
    \begin{tabular}{ lllllll }
        \toprule
        iauname &     ra &   dec & bar\_total-votes & bar\_weak & bar\_weak\_fraction & bar\_weak\_debiased \\
        \midrule
         J112953.88-000427.4 & 172.47 & -0.07 &               16 &         1 &                 0.06 &                 0.15 \\
         J104325.29+190335.0 & 160.86 & 19.06 &                2 &         0 &                 0.00 &                 0.00 \\
         J104629.54+115415.1 & 161.62 & 11.90 &                4 &         2 &                 0.50 &                 - \\
         J082950.68+125621.8 & 127.46 & 12.94 &                0 &         0 &                    - &                    - \\
         J122056.00-015022.0 & 185.23 & -1.84 &                3 &         0 &                 0.00 &                 - \\
        \bottomrule
    \end{tabular}
    \caption{Sample of GZD-5 volunteer classifications, with illustrative subset of columns. Columns: `iauname' galaxy identifier from NASA-Sloan Atlas; RA and Dec, similarly; `Bar' question total votes for all answers; `Bar' question votes for `Weak' answer; fraction of `Bar' question votes for `Weak' answer; estimated fraction after applying redshift debiasing (Sec. \ref{sec:redshift_debiasing}). Other questions and answers follow the same pattern (not shown for brevity). Full schema online.}
    \label{tab:sample_volunteers}
\end{table*}

\begin{table*}
    \begin{tabular}{llllll}
        \toprule
                     iauname &     RA &   Dec & bar\_proportion\_asked &                           bar\_weak\_concentrations & bar\_weak\_fraction \\
        \midrule
         J112953.88-000427.4 & 172.47 & -0.07 &                   0.14 &  [6.158, 5.0723, 5.4842, ... &                     0.09 \\
         J104325.29+190335.0 & 160.86 & 19.06 &                   0.13 &  [4.3723, 4.5933, 4.8582... &                     0.07 \\
         J100927.56+071112.4 & 152.36 &  7.19 &                   0.58 &  [9.3129, 10.3911, 8.4791... &                     0.40 \\
         J143254.45+034938.1 & 218.23 &  3.83 &                   0.55 &  [13.2981, 12.2639, 8.8957... &                     0.26 \\
         J135942.73+010637.3 & 209.93 &  1.11 &                   0.77 &  [15.6247, 15.6893, 14.72.... &                     0.28 \\
        \bottomrule
    \end{tabular}
    \caption{Sample of automated classifications (GZD-5 schema), with illustrative subset of columns. Columns: `iauname' galaxy identifier from NASA-Sloan Atlas; RA and Dec, similarly; proportion of volunteers estimated to be asked the `Bar' question (i.e. the product of the preceding vote fractions) for estimating relevance; Dirichlet concentrations defining the predicted posterior for the `Bar' question and `Weak' answer (see Sec. \ref{sec:automated}); predicted fraction of `Bar' question votes for the `Weak' answer derived from those concentrations. Other questions and answers follow the same pattern (not shown for brevity). Full schema online.}
    \label{tab:sample_ml}
\end{table*}

\section*{Acknowledgements}
\label{sec:acknowledgements}

The data in this paper are the result of the efforts of the Galaxy Zoo volunteers, without whom none of this work would be possible. Their efforts are individually acknowledged at \href{http://authors.galaxyzoo.org}{http://authors.galaxyzoo.org}. We would also like to thank our volunteer translators; Mei-Yin Chou, Antonia Fernández Figueroa, Rodrigo Freitas, Na'ama Hallakoun, Lauren Huang, Alvaro Menduina, Beatriz Mingo, Verónica Motta, João Retrê, and Erik Rosenberg.

We would like to thank Dustin Lang for creating the \href{wwww.legacysurvey.org}{legacysurvey.org} cutout service and for contributing image processing code. We also thank Sugata Kaviraj and Matthew Hopkins for helpful discussions.

MW acknowledges funding from the Science and Technology Funding Council (STFC) Grant Code ST/R505006/1.
We also acknowledge support from STFC under grant ST/N003179/1. 

RJS acknowledges funding from Christ Church, University of Oxford. 

LF acknowledges partial support from US National Science Foundation award OAC 1835530; VM and LF acknowledge partial support from NSF AST 1716602.

This publication uses data generated via the Zooniverse.org platform, development of which is funded by generous support, including a Global Impact Award from Google, and by a grant from the Alfred P. Sloan Foundation.

This research made use of the open-source Python scientific computing ecosystem, including SciPy \citep{Jones2001}, Matplotlib \citep{Hunter2007}, scikit-learn \citep{Pedregosa2011}, scikit-image \citep{VanderWalt2014} and Pandas \citep{McKinney2010}.

This research made use of Astropy, a community-developed core Python package for Astronomy \citep{TheAstropyCollaboration2018}.

This research made use of TensorFlow \citep{Abadi2015}.

The Legacy Surveys consist of three individual and complementary projects: the Dark Energy Camera Legacy Survey (DECaLS; NSF's OIR Lab Proposal ID \# 2014B-0404; PIs: David Schlegel and Arjun Dey), the Beijing-Arizona Sky Survey (BASS; NSF's OIR Lab Proposal ID \# 2015A-0801; PIs: Zhou Xu and Xiaohui Fan), and the Mayall z-band Legacy Survey (MzLS; NSF's OIR Lab Proposal ID \# 2016A-0453; PI: Arjun Dey). DECaLS, BASS and MzLS together include data obtained, respectively, at the Blanco telescope, Cerro Tololo Inter-American Observatory, The NSF's National Optical-Infrared Astronomy Research Laboratory (NSF's OIR Lab); the Bok telescope, Steward Observatory, University of Arizona; and the Mayall telescope, Kitt Peak National Observatory, NSF's OIR Lab. The Legacy Surveys project is honored to be permitted to conduct astronomical research on Iolkam Du'ag (Kitt Peak), a mountain with particular significance to the Tohono O'odham Nation.

The NSF's OIR Lab is operated by the Association of Universities for Research in Astronomy (AURA) under a cooperative agreement with the National Science Foundation.

This project used data obtained with the Dark Energy Camera (DECam), which was constructed by the Dark Energy Survey (DES) collaboration. Funding for the DES Projects has been provided by the U.S. Department of Energy, the U.S. National Science Foundation, the Ministry of Science and Education of Spain, the Science and Technology Facilities Council of the United Kingdom, the Higher Education Funding Council for England, the National Center for Supercomputing Applications at the University of Illinois at Urbana-Champaign, the Kavli Institute of Cosmological Physics at the University of Chicago, Center for Cosmology and Astro-Particle Physics at the Ohio State University, the Mitchell Institute for Fundamental Physics and Astronomy at Texas A\&M University, Financiadora de Estudos e Projetos, Fundacao Carlos Chagas Filho de Amparo, Financiadora de Estudos e Projetos, Fundacao Carlos Chagas Filho de Amparo a Pesquisa do Estado do Rio de Janeiro, Conselho Nacional de Desenvolvimento Cientifico e Tecnologico and the Ministerio da Ciencia, Tecnologia e Inovacao, the Deutsche Forschungsgemeinschaft and the Collaborating Institutions in the Dark Energy Survey. The Collaborating Institutions are Argonne National Laboratory, the University of California at Santa Cruz, the University of Cambridge, Centro de Investigaciones Energeticas, Medioambientales y Tecnologicas-Madrid, the University of Chicago, University College London, the DES-Brazil Consortium, the University of Edinburgh, the Eidgenossische Technische Hochschule (ETH) Zurich, Fermi National Accelerator Laboratory, the University of Illinois at Urbana-Champaign, the Institut de Ciencies de l'Espai (IEEC/CSIC), the Institut de Fisica d'Altes Energies, Lawrence Berkeley National Laboratory, the Ludwig-Maximilians Universitat Munchen and the associated Excellence Cluster Universe, the University of Michigan, the National Optical Astronomy Observatory, the University of Nottingham, the Ohio State University, the University of Pennsylvania, the University of Portsmouth, SLAC National Accelerator Laboratory, Stanford University, the University of Sussex, and Texas A\&M University.

BASS is a key project of the Telescope Access Program (TAP), which has been funded by the National Astronomical Observatories of China, the Chinese Academy of Sciences (the Strategic Priority Research Program `The Emergence of Cosmological Structures' Grant \# XDB09000000), and the Special Fund for Astronomy from the Ministry of Finance. The BASS is also supported by the External Cooperation Program of Chinese Academy of Sciences (Grant \# 114A11KYSB20160057), and Chinese National Natural Science Foundation (Grant \# 11433005).

The Legacy Survey team makes use of data products from the Near-Earth Object Wide-field Infrared Survey Explorer (NEOWISE), which is a project of the Jet Propulsion Laboratory/California Institute of Technology. NEOWISE is funded by the National Aeronautics and Space Administration.

The Legacy Surveys imaging of the DESI footprint is supported by the Director, Office of Science, Office of High Energy Physics of the U.S. Department of Energy under Contract No. DE-AC02-05CH1123, by the National Energy Research Scientific Computing Center, a DOE Office of Science User Facility under the same contract; and by the U.S. National Science Foundation, Division of Astronomical Sciences under Contract No. AST-0950945 to NOAO.

\section*{Data Availability}

The data underlying this article are available via Zenodo at \href{https://doi.org/10.5281/zenodo.4573248}{https://doi.org/10.5281/zenodo.4573248}. Any future data updates will be released using Zenodo versioning - please check you are viewing the latest version. The code underlying this article is available at \href{https://github.com/mwalmsley/zoobot}{https://github.com/mwalmsley/zoobot}.



\bibliographystyle{mnras}
\bibliography{bibliography}


\bsp	
\label{lastpage}
\end{document}


\appendix
\section{Galaxies with Confident Automated Classifications}

\begin{figure*}
    \centering
    \includegraphics[height=0.7\textheight,trim=140 160 140 150, clip]{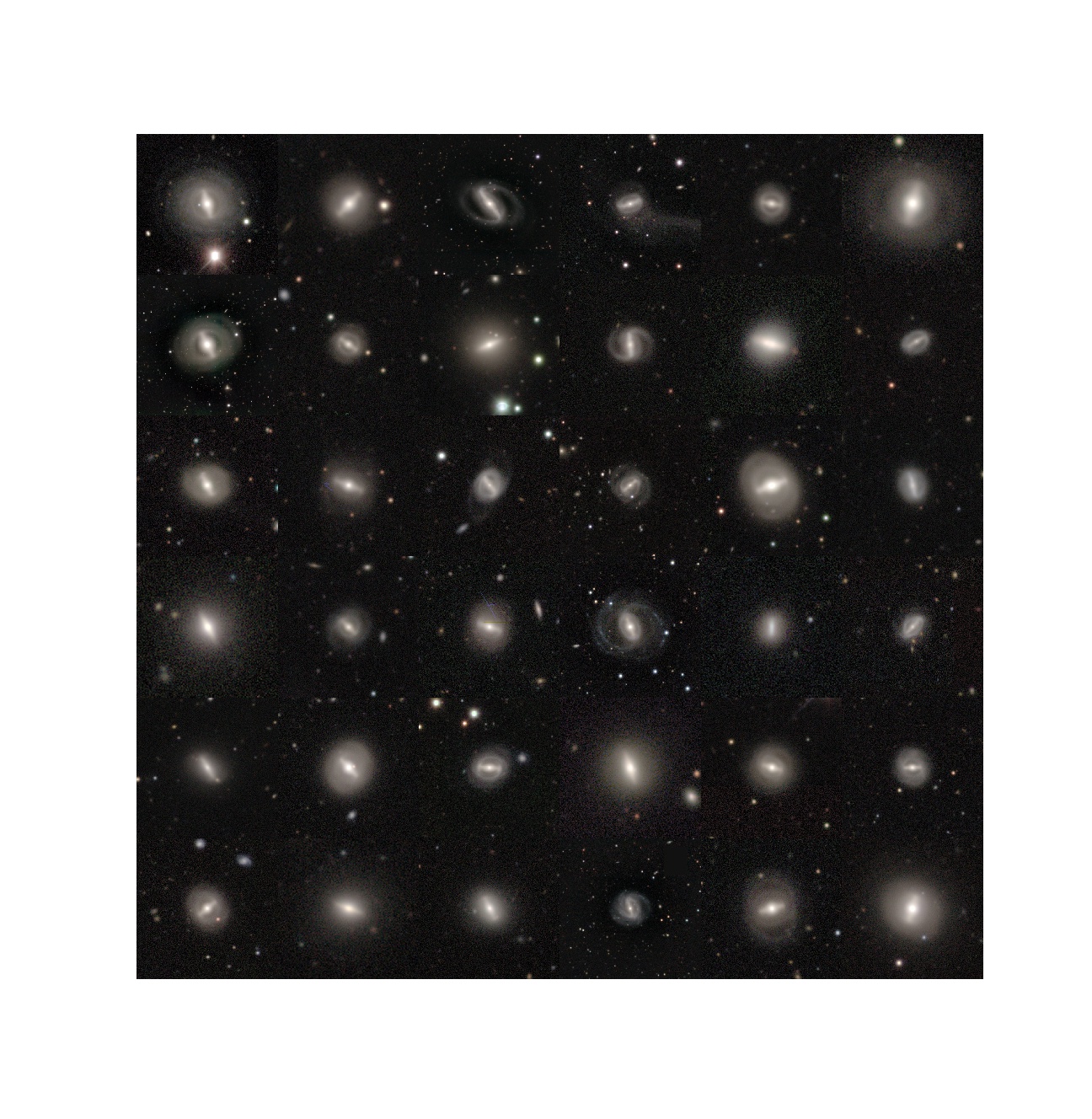}
    \caption{Galaxies automatically classified as most likely (highest mean posterior) to be strongly barred.}
    \label{fig:bar_strong_grid}
\end{figure*}

\begin{figure*}
    \centering
    \includegraphics[height=0.7\textheight,trim=140 160 140 150, clip]{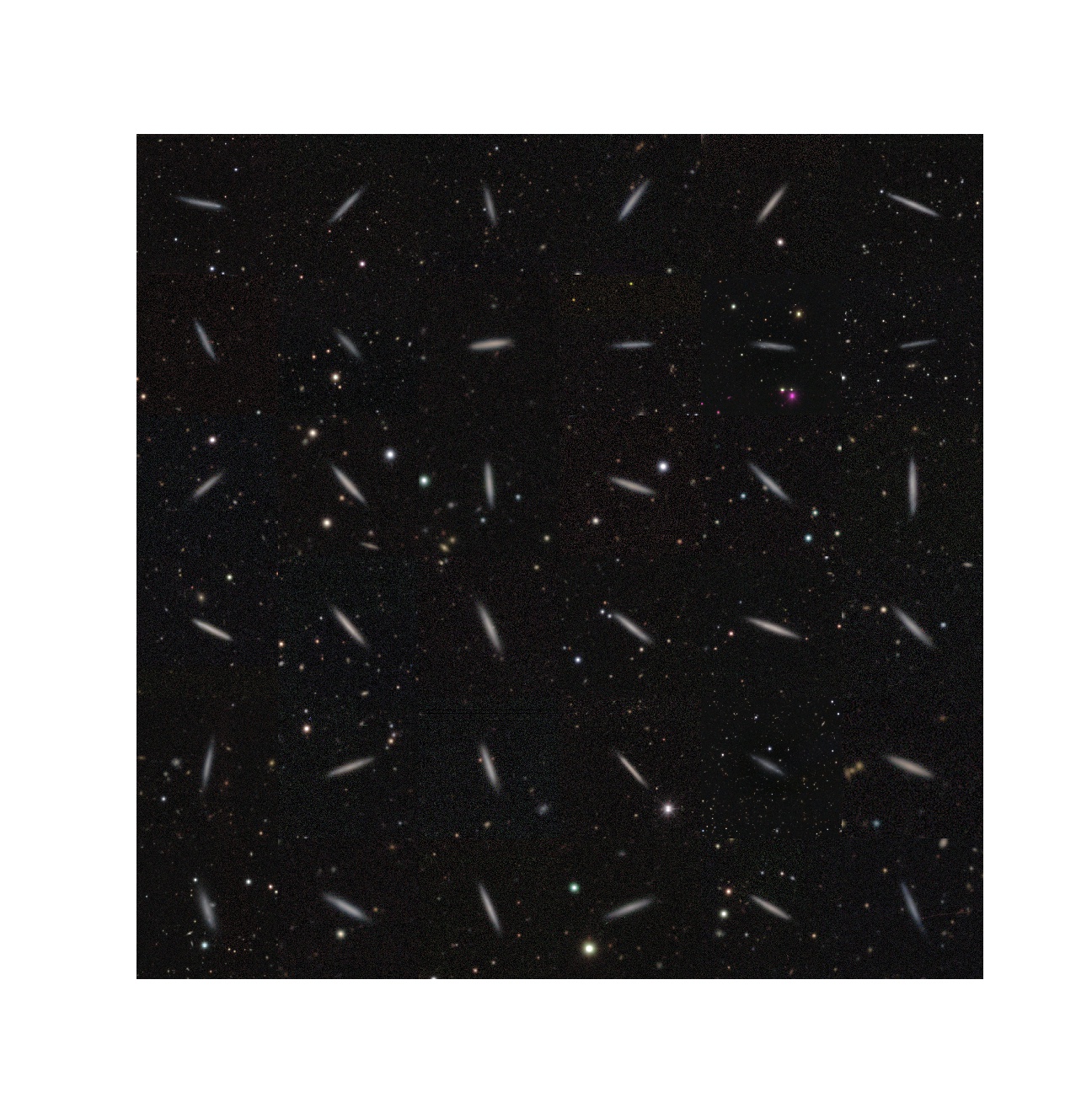}
    \caption{Galaxies automatically classified as most likely (highest mean posterior) to be edge-on with no bulge.}
    \label{fig:edge_on_no_bulge_grid}
\end{figure*}

\begin{figure*}
    \centering
    \includegraphics[height=0.7\textheight,trim=140 160 140 150, clip]{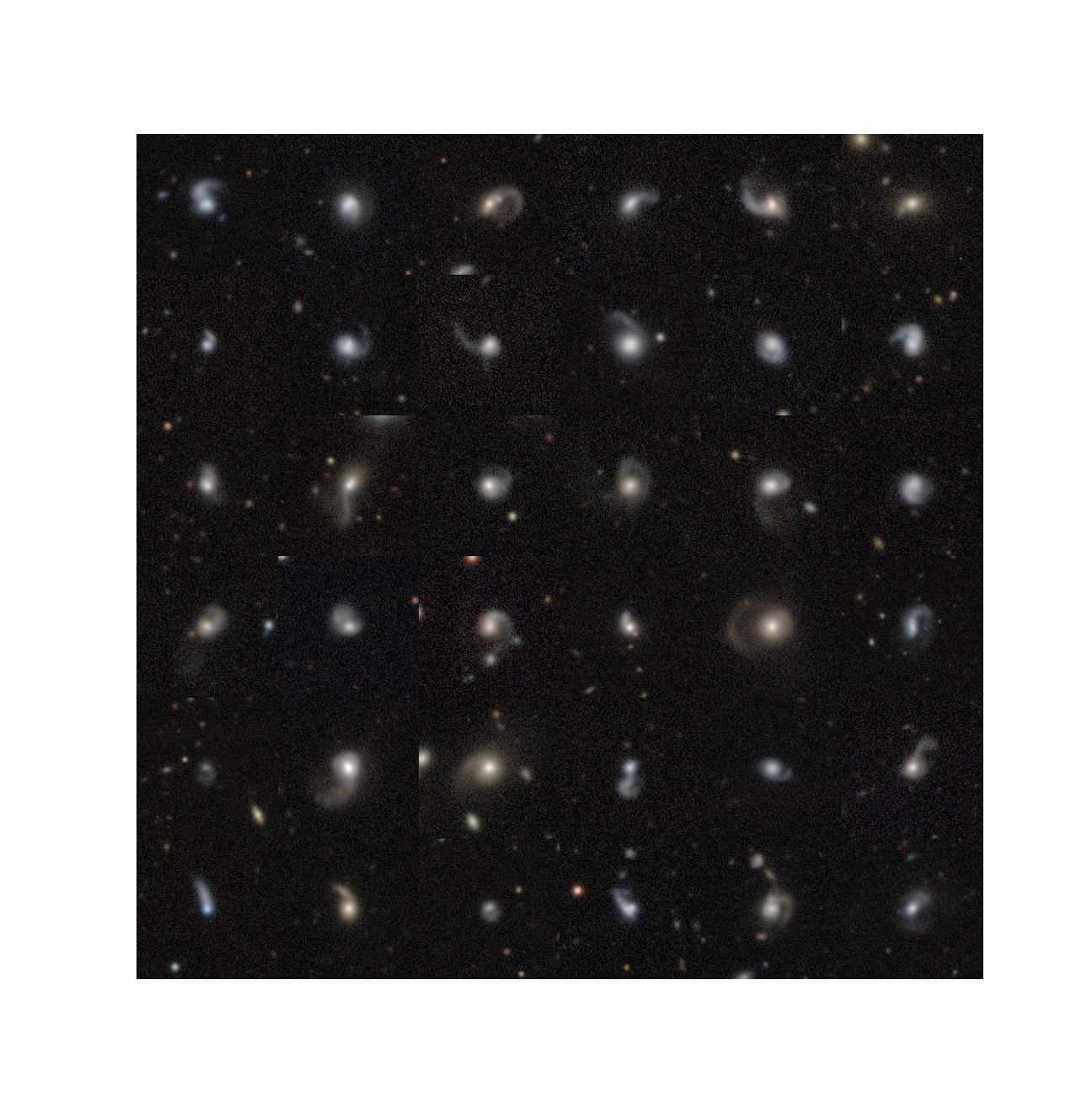}
    \caption{Galaxies automatically classified as most likely (highest mean posterior) to have exactly one spiral arm.}
    \label{fig:spiral_1_grid}
\end{figure*}

\begin{figure*}
    \centering
    \includegraphics[height=0.7\textheight,trim=140 160 140 150, clip]{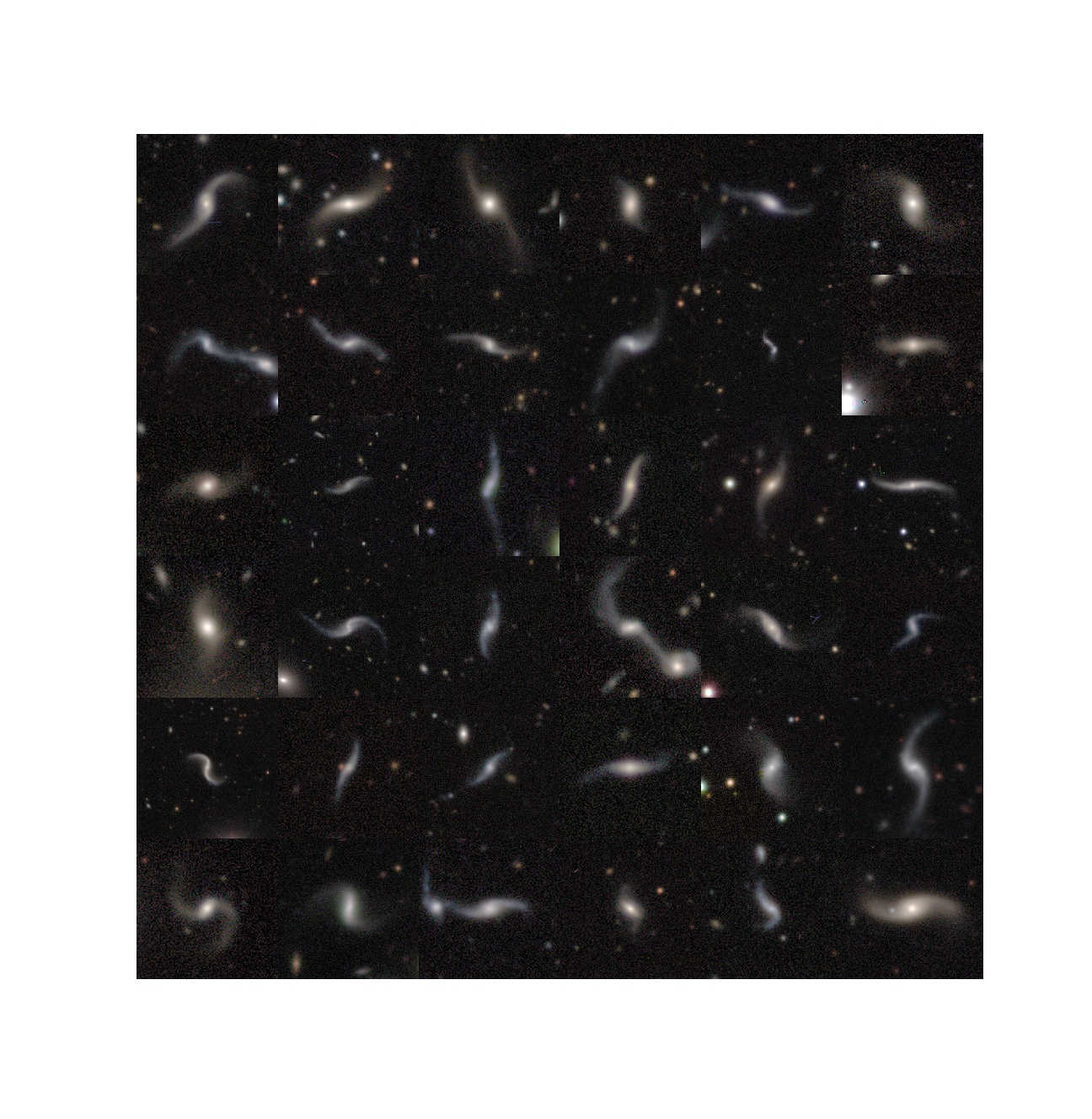}
    \caption{Galaxies automatically classified as most likely (highest mean posterior) to have loosely wound spiral arms.}
    \label{fig:winding_loose_grid}
\end{figure*}

\begin{figure*}
    \centering
    \includegraphics[height=0.7\textheight,trim=140 160 140 150, clip]{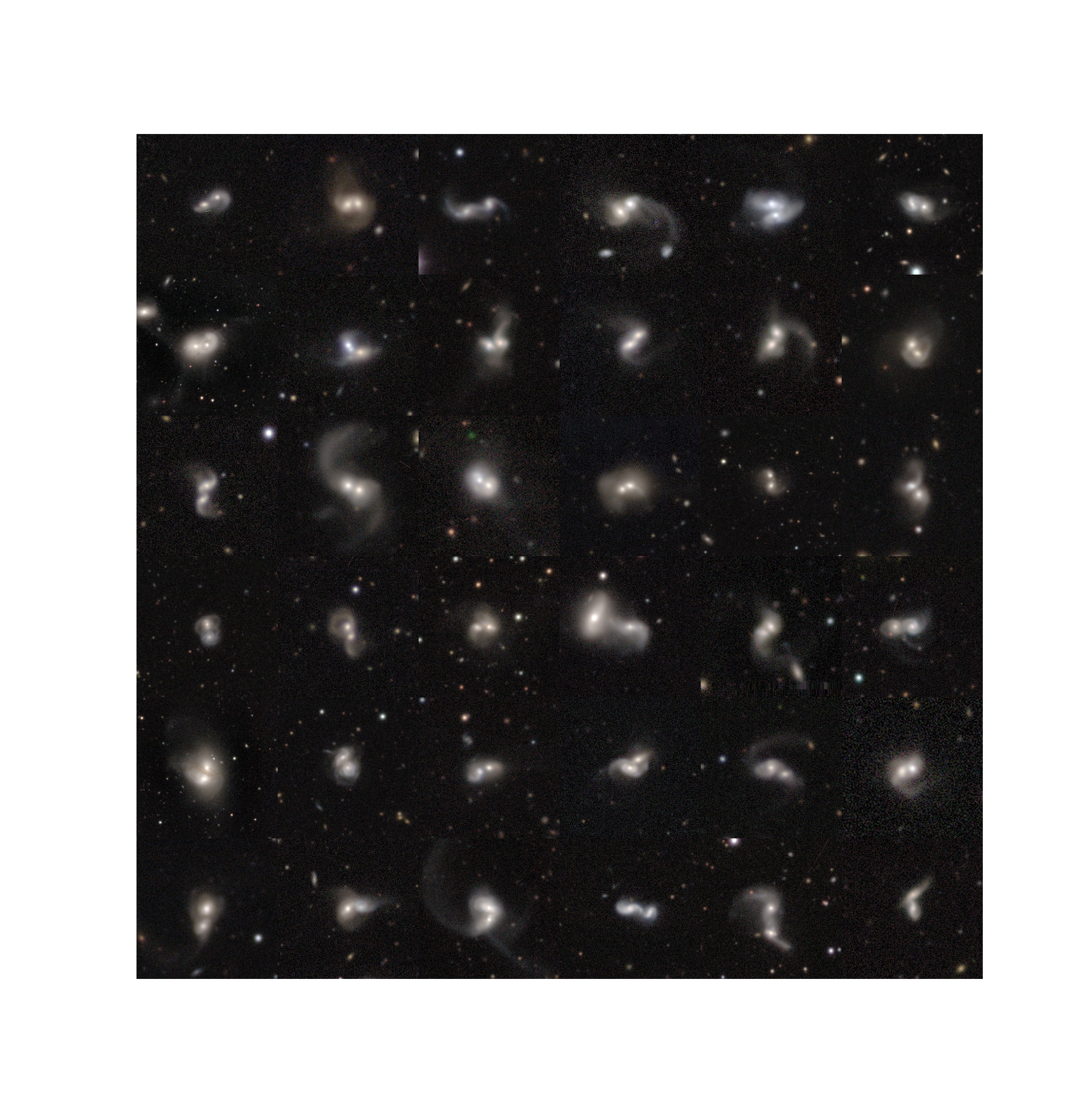}
    \caption{Galaxies automatically classified as most likely (highest mean posterior) to be mergers, with automatic `featured' vote fraction > 0.5. Only one thumbnail per galaxy pair is shown.}
    \label{fig:merger_grid}
\end{figure*}

To intuitively demonstrate the performance of our automated classifier, we show, for a selection of detailed morphology questions, the galaxies with the most confident automated classifications for that question. We show the galaxies with the highest mean posterior for being strongly barred (Fig. \ref{fig:bar_strong_grid}), edge-on and bulgeless (Fig \ref{fig:edge_on_no_bulge_grid}), one-armed spirals (Fig. \ref{fig:spiral_1_grid}), loosely wound spirals (Fig. \ref{fig:winding_loose_grid}) and mergers (Fig. \ref{fig:merger_grid}). We present the galaxies here as shown to Galaxy Zoo volunteers (in color and at 424x424 pixel resolution), but the model makes predictions on more challenging greyscale 224x224 pixel images.

